\documentclass[traditabstract,longauth]{aa}

\usepackage{graphicx}
\usepackage{txfonts}
\usepackage{natbib}
\usepackage[colorlinks=true,citecolor=blue]{hyperref}
\usepackage{xspace}
\usepackage{longtable}
\usepackage{array}
\usepackage{lscape}
\usepackage{rotating}
\usepackage[labelformat=simple]{caption,subcaption}
\usepackage{ifthen}

\bibpunct{(}{)}{;}{a}{}{,}

\newcommand{\MJup}{\ensuremath{M_{\mathrm{Jup}}}\xspace}

\newcommand{\MSun}{\ensuremath{M_{\odot}}\xspace}

\newcommand{\MEarth}{\ensuremath{M_{\oplus}}\xspace}

\newcommand{\as}{\hbox{$^{\prime\prime}$}\xspace}

\newcommand{\surv}[1]
{
  \ifthenelse{\equal{#1}{M05}}{\citetalias{masciadri2005}}{}%
  \ifthenelse{\equal{#1}{L05}}{\citetalias{lowrance2005}}{}%
  \ifthenelse{\equal{#1}{S06}}{S06}{}%
  \ifthenelse{\equal{#1}{K07}}{\citetalias{kasper2007}}{}%
  \ifthenelse{\equal{#1}{B07}}{\citetalias{biller2007}}{}%
  \ifthenelse{\equal{#1}{L07}}{\citetalias{lafreniere2007}}{}%
  \ifthenelse{\equal{#1}{C10}}{\citetalias{chauvin2010}}{}%
  \ifthenelse{\equal{#1}{H10}}{\citetalias{heinze2010}}{}%
  \ifthenelse{\equal{#1}{V12}}{\citetalias{vigan2012}}{}%
  \ifthenelse{\equal{#1}{R13}}{\citetalias{rameau2013}}{}%
  \ifthenelse{\equal{#1}{B13}}{\citetalias{biller2013}}{}%
  \ifthenelse{\equal{#1}{B14}}{\citetalias{brandt2014}}{}%
  \ifthenelse{\equal{#1}{C15}}{\citetalias{chauvin2015}}{}%
}

\defcitealias{masciadri2005}{M05}
\defcitealias{lowrance2005}{L05}
\defcitealias{kasper2007}{K07}
\defcitealias{biller2007}{B07}
\defcitealias{lafreniere2007}{L07}
\defcitealias{chauvin2010}{C10}
\defcitealias{heinze2010}{H10}
\defcitealias{vigan2012}{V12}
\defcitealias{rameau2013}{R13}
\defcitealias{biller2013}{B13}
\defcitealias{brandt2014}{B14}
\defcitealias{chauvin2015}{C15}

\begin{document}

\title{The VLT/NaCo large program to probe the occurrence of exoplanets and brown dwarfs at wide orbits\thanks{Based on observations collected at the European Southern Observatory, Chile (ESO Large Program 184.C-0157 and Open Time 089.C-0137A and 090.C-0252A).}}
\subtitle{IV. Gravitational instability rarely forms wide, giant planets}
\titlerunning{The VLT/NaCo large program to probe the occurrence of exoplanets and brown dwarfs at wide orbits. IV.}

\author{
    A. Vigan\inst{1} \and 
    M. Bonavita\inst{2,3} \and 
    B. Biller\inst{2,4} \and 
    D. Forgan\inst{5} \and 
    K. Rice\inst{2} \and 
    G. Chauvin\inst{7,8} \and 
    S. Desidera\inst{3} \and
    J.-C. Meunier\inst{1} \and
    P. Delorme\inst{7,8} \and
    J. E. Schlieder\inst{9,4} \and
    M. Bonnefoy\inst{7,8} \and
    J. Carson\inst{10,4} \and
    E. Covino\inst{11} \and
    J. Hagelberg\inst{7,8} \and
    T. Henning\inst{4} \and
    M. Janson\inst{12,4} \and
    A.-M. Lagrange\inst{7,8} \and
    S. P. Quanz\inst{13} \and
    A. Zurlo\inst{14,15,1} \and
    J.-L. Beuzit\inst{7,8} \and
    A. Boccaletti\inst{16} \and
    E. Buenzli\inst{13} \and
    M. Feldt\inst{4} \and
    J. H. V. Girard\inst{17} \and
    R. Gratton\inst{3} \and
    M. Kasper\inst{18} \and
    H. Le Coroller\inst{1} \and
    D. Mesa\inst{3} \and
    S. Messina\inst{19} \and
    M. Meyer\inst{13} \and
    G. Montagnier\inst{20} \and
    C. Mordasini\inst{21} \and
    D. Mouillet\inst{7,8} \and 
    C. Moutou\inst{22,1} \and
    M. Reggiani\inst{13} \and
    D. Segransan\inst{23} \and
    C. Thalmann\inst{13}
}

\institute{
    Aix Marseille Univ, CNRS, LAM, Laboratoire d'Astrophysique de Marseille, Marseille, France \\ \email{\href{mailto:arthur.vigan@lam.fr}{arthur.vigan@lam.fr}} 
    \and
    SUPA, Institute for Astronomy, The University of Edinburgh, Royal Observatory, Blackford Hill, Edinburgh, EH9 3HJ, UK 
    \and
    INAF - Osservatorio Astronomico di Padova, Vicolo dell'Osservatorio 5, I-35122, Padova, Italy 
    \and    
    Max-Planck Institute for Astronomy, K\"onigstuhl 17, 69117 Heidelberg, Germany 
    \and 
    SUPA, School of Physics and Astronomy, University of St Andrews, North Haugh, St Andrews KY16 9SS, UK 
    \and
    St Andrews Centre for Exoplanet Science 
    \and
    Universit\'e Grenoble Alpes, IPAG, F-38000 Grenoble, France  
    \and
    CNRS, IPAG, F-38000 Grenoble, France 
    \and
    NASA Exoplanet Science Institute, California Institute of Technology, 770 S. Wilson Ave., Pasadena, CA, USA 
    \and
    Department of Physics and Astronomy, College of Charleston, Charleston, SC 29424 , USA 
    \and
    INAF Osservatorio Astronomico di Capodimonte via Moiarello 16, 80131 Napoli, Italy 
    \and
    Department of Astronomy, Stockholm University, AlbaNova University Center, 106 91 Stockholm, Sweden 
    \and
    Institute for Astronomy, ETH Zurich, Wolfgang-Pauli-Strasse 27, 8093 Zurich, Switzerland 
    \and
    N\'ucleo de Astronom\'ia, Facultad de Ingenier\'ia, Universidad Diego Portales, Av. Ejercito 441, Santiago, Chile 
    \and
    Millennium Nucleus ``Protoplanetary Disk'', Departamento de Astronom\'ia, Universidad de Chile, Casilla 36-D, Santiago, Chile 
    \and
    LESIA, Observatoire de Paris, CNRS, Universit\'e Paris Diderot, Universit\'e Pierre et Marie Curie, 5 place Jules Janssen, 92190 Meudon, France 
    \and
    European Southern Observatory, Alonso de Cordova 3107, Vitacura, Santiago, Chile 
    \and
    European Southern Observatory, Karl-Schwarzschild-Str. 2, 85748 Garching, Germany 
    \and
    INAF – Catania Astrophysical Observatory, via S. So a 78, 95123 Catania, Italy 
    \and
    Observatoire de Haute-Provence, CNRS, Universit\'e d'Aix-Marseille, 04870 Saint-Michel-l'Observatoire, France 
    \and
    Physikalisches Institut, University of Bern, Sidlerstrasse 5, 3012 Bern, Switzerland 
    \and
    CNRS, CFHT, 65-1238 Mamalahoa Hwy, Kamuela HI, USA 
    \and
    Observatoire Astronomique de l'Universit\'e de Gen\`eve, Chemin des Maillettes 51, 1290 Sauverny, Switzerland 
}

\date{Received 25 November 2016; accepted 6 March 2017}

\abstract{
  Understanding the formation and evolution of giant planets ($\ge$1~\MJup) at wide orbital separation ($\ge$5~AU) is one of the goals of direct imaging. Over the past 15 years, many surveys have placed strong constraints on the occurrence rate of wide-orbit giants, mostly based on non-detections, but very few have tried to make a direct link with planet formation theories. In the present work, we combine the results of our previously published VLT/NaCo large program with the results of 12 past imaging surveys to constitute a statistical sample of 199 FGK stars within 100~pc, including three stars with sub-stellar companions. Using Monte Carlo simulations and assuming linear flat distributions for the mass and semi-major axis of planets, we estimate the sub-stellar companion frequency to be within 0.75--5.70\% at the 68\% confidence level (CL) within 20--300~AU and 0.5--75~\MJup, which is compatible with previously published results. We also compare our results with the predictions of state-of-the-art population synthesis models based on the gravitational instability (GI) formation scenario by \citet{forgan2013}, with and without scattering. We estimate that in both the scattered and non-scattered populations, we would be able to detect more than 30\% of companions in the 1--75~\MJup range (95\% CL). With the three sub-stellar detections in our sample, we estimate the fraction of stars that host a planetary system formed by GI to be within 1.0--8.6\% (95\% CL). We also conclude that even though GI is not common, it predicts a mass distribution of wide-orbit massive companions that is much closer to what is observed than what the core accretion scenario predicts. Finally, we associate the present paper with the release of the Direct Imaging Virtual Archive (DIVA, \url{http://cesam.lam.fr/diva/}), a public database that aims at gathering the results of past, present, and future direct imaging surveys.
}


\maketitle

\section{Introduction}
\label{sec:introduction}

Among the various exoplanet detection methods, direct imaging is the best suited to look for sub-stellar companions orbiting at large orbital separation (>10~AUs) around nearby stars (<100~pc). Over the past 15 years, multiple surveys targeting a variety of stars with different spectral type, distance, age, or metallicity have been successful at placing tight constraints on the frequency of giant planets and brown dwarfs in the 50--1000~AU range (see \citet{bowler2016} for a recent review). Despite the development of optimized observing strategies and data analysis methods, only a limited number of sub-stellar companions have been detected, leading to the conclusion that the frequency of these objects is very low (typically $\lesssim$5\%; \citealt{galicher2016}). At closer orbital separations, in the 5--50~AU range, the question remains largely unanswered, although initial results \citep{macintosh2015,wagner2016} from on-going large-scale direct imaging surveys with the new generation of high-contrast imagers and spectrographs \citep{beuzit2008,macintosh2014,guyon2010} suggest that the frequency of young giant planets is within a few percent. 

The few known directly imaged giant planets as well as the overall paucity of this type of object at large separations raise the question of planetary formation. The two competing mainstream scenarios are core accretion (CA) and gravitational instability\footnote{Sometimes also referred to as disk instability.} (GI). In the CA model, giant planets are formed in a multi-stage process entailing the buildup of a 10--15~\MEarth core followed by rapid accretion of gas from the protoplanetary disk \citep{Pollack1996,alibert2004}. This model has been very powerful in explaining a lot of the properties of the exoplanet population detected within 5~AU, but it faces major difficulties in explaining the formation of giant planets farther out than $\sim$20~AU because of the timescales involved \citep{alibert2005,kennedy2008}. Pebble accretion has been proposed as a potential solution to this issue \citep{lambrechts2012,levison2015}, but the most recent simulations show that this mechanism does not really form giant planets (several \MJup) at large orbital distances \citep[][their Figs. 4 and 5]{bitsch2015}.

In the GI model, the planets are the result of gravitational fragmentation and collapse of clumps in the disk \citep{boss1998,vorobyov2013}. This model could more readily explain the existence of gas giants very far from their star, but the conditions that can lead to disk fragmentation are still not fully understood \citep{meru2011,Paardekooper2012, rice2012,Rice2014,Young2016}. In any case, disk-planet interactions \citep{kley2012} or planet-planet scattering \citep{veras2009,dawson2013} will likely impact on the original semi-major axis distribution of planets, resulting in exoplanets covering a wide range of possible masses, sizes, locations, and compositions.

Observational constraints are crucial in validating or invalidating planetary formation models, and in investigating the migration processes at play during the early evolution of planetary systems. In this regard, planet population synthesis is an interesting approach to statistically compare observations with theory and detect potential shortcomings in the modeling (see the review by \citealt{benz2014}). One of the important early successes of planet population synthesis has been to be able to provide a consistent picture of the mass versus semi-major axis diagram for planets up to a few AUs \citep{ida2004,mordasini2009}. 

For directly imaged planets, population synthesis has so far mainly been used for comparison with individual objects, like the giant planet around $\beta$~Pictoris \citep{bonnefoy2013}, or as an illustration of the potential for future exoplanet imagers \citep{rameau2013}. One of the reasons is that population synthesis developments have been primarily based on the CA model, which forms very few massive planets at large separations, therefore making any comparison with results from past direct imaging surveys pointless. Although planet-planet scattering can send planets much farther out in the systems and allow their direct detection, population synthesis models based on CA only recently started to incorporate several planet embryos per disk and their subsequent dynamical evolution \citep{alibert2013,ida2013}. Predictions are currently too premature for comparison with past direct imaging surveys, but will be a key aspect for the forthcoming analysis of the large surveys on-going with VLT/SPHERE \citep[Spectro-Polarimetric High contrast Exoplanet REsearch;][]{beuzit2008} and Gemini/GPI \citep[Gemini Planet Imager;][]{macintosh2014}.

Comparison between direct imaging observations and predictions of the GI theory has only been explored statistically in two surveys \citep{janson2011,janson2012,rameau2013}. They used physical considerations to define the domains in the (mass, semi-major axis) parameter space where formation by GI would be allowed, and used the non-detection in their surveys to set upper limits on the giant planet frequency. These studies, however, do not consider the subsequent evolution of a population that may form via GI. Planet population synthesis based on the GI theory has taken much longer than CA to mature, and only lately has a consistent model including multi-object systems been formulated \citep{forgan2013,forgan2015}. This model predicts the formation of a wide range of sub-stellar objects, a large portion of which could be detected by direct imaging with conventional or extreme adaptive optics systems.

In the present paper, we put the \citet{forgan2013} population synthesis model to the test by comparing its outcome to direct imaging results obtained around 200 young, solar-type (FGK) stars. In Sect.~\ref{sec:sample_description} we describe our full statistical sample based on 13 previously published direct imaging surveys spanning more than ten years. In Sect.~\ref{sec:detection_limits}, we discuss the detection limits and we present the Direct Imaging Virtual Archive (DIVA), a public database associated with this paper. In Sect.~\ref{sec:planet_population_synthesis_models} we detail the population synthesis and evolutionary models considered in our analysis and then in Sect.~\ref{sec:statistical_analysis} we present our statistical analysis based on outputs of population synthesis. Finally, we discuss our results and present our conclusions in Sect.~\ref{sec:conclusions}.

\section{Sample description}
\label{sec:sample_description}

\begin{figure*}
  \centering
  \includegraphics[width=1\textwidth]{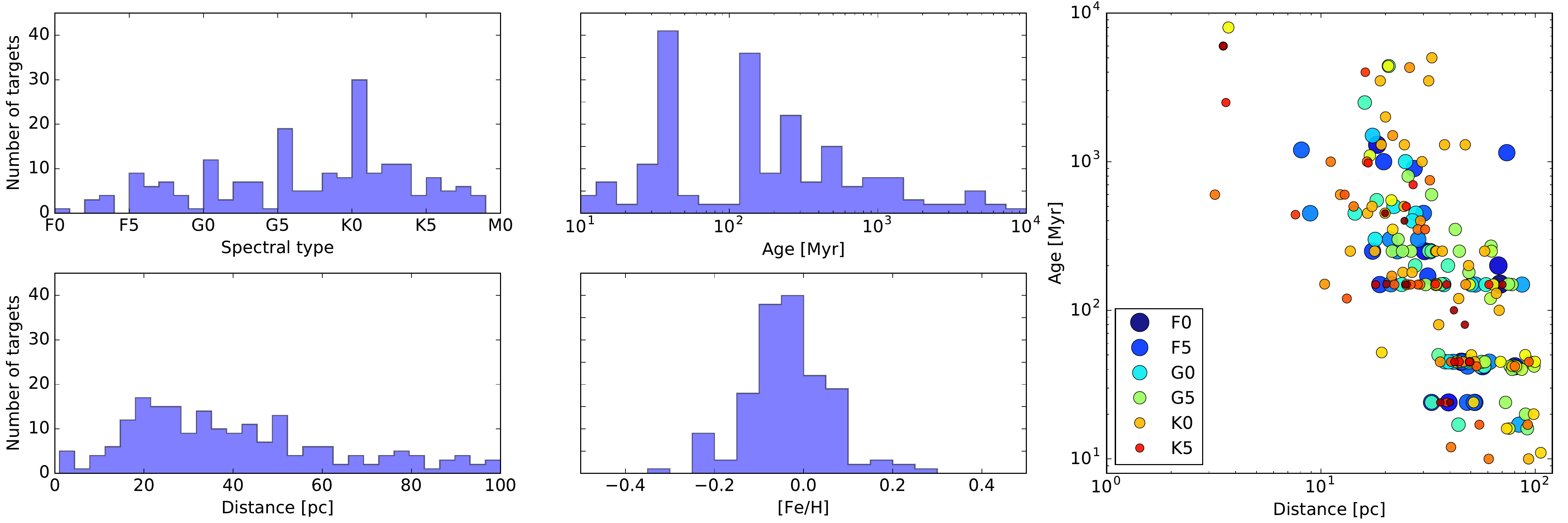}
  \caption{Histograms summarising the properties of all the targets considered in the final sample: spectral type, age, distance, and metallicity (from left to right). The complete sample includes 199 individuals targets. All the properties are detailed in Appendix~\ref{sec:full_sample}.}
  \label{fig:sample_histograms}
\end{figure*}

\begin{table*}
  \caption{Surveys considered for the construction of the NACO-LP archive sample.}
  \label{tab:surveys}
  \centering
  \begin{tabular}{lccccccl}
  \hline\hline
  Name           & Telescope   & Instrument & Techniques\tablefootmark{a} & Band & $N_{\rm survey}$\tablefootmark{b}  & $N_{\rm NaCo-LP}$ & Reference \\
  \hline
  \surv{M05}     & VLT         & NaCo       & DI        & $H, K$ & 18       & 5      & \citet{masciadri2005} \\
  \surv{L05}     & HST         & NICMOS     & COR       & $H$    & 45       & 17     & \citet{lowrance2005} \\
  \surv{S06}     & HST         & NICMOS     & ADI       & $H$    & 116      & 47     & Song et al. (unpublished)\tablefootmark{c} \\
  \surv{K07}     & VLT         & NaCo       & ADI, RDI  & $L$    & 22       & 6      & \citet{kasper2007} \\
  \surv{B07}     & VLT         & NaCo       & SDI       & $H$    & 45       & 24     & \citet{biller2007} \\
  \surv{L07}     & Gemini      & NIRI       & ADI       & $H$    & 85       & 50     & \citet{lafreniere2007} \\
  \surv{C10}     & VLT         & NaCo       & DI        & $H, K$ & 91       & 29     & \citet{chauvin2010} \\
  \surv{H10}     & MMT         & Clio       & ADI       & $L, M$ & 54       & 27     & \citet{heinze2010} \\
  \surv{V12}     & Gemini, VLT & NIRI, NaCo & ADI       & $H, K$ & 42       & 4      & \citet{vigan2012} \\
  \surv{R13}     & VLT         & NaCo       & ADI       & $L$    & 59       & 21     & \citet{rameau2013} \\
  \surv{B13}     & Gemini      & NICI       & ASDI      & $H$    & 80       & 22     & \citet{biller2013} \\
  \surv{B14}     & Subaru      & HiCIAO     & ADI       & $H$    & 63       & 29     & \citet{brandt2014} \\
  \surv{C15}     & VLT         & NaCo       & ADI       & $H$    & 85       & 47     & \citet{chauvin2015} \\
  \hline
  \end{tabular}
  \tablefoot{\tablefoottext{a}{Techniques: ADI = angular differential imaging, usually coupled with saturated imaging; ASDI = angular and spectral differential imaging; COR = coronagraphy; DI = direct imaging (saturated); RDI = reference differential imaging; SDI = spectral differential imaging.} \tablefoottext{b}{The sum of $N_{\rm survey}$ for all surveys is larger than 199, the total number of stars considered in our analysis, because some stars were observed by several surveys.} \tablefoottext{c}{HST, cycle 13, \href{https://archive.stsci.edu/proposal_search.php?id=10176&mission=hst}{GO10176}}}
\end{table*}

The current work was started in the context of the VLT/NaCo large program to probe the occurrence of exoplanets and brown dwarfs at wide orbits (ESO program 184.C-0157, P.I. J.-L. Beuzit), hereafter NaCo-LP. The general philosophy of the NaCo-LP sample selection has previously been described in \citet{desidera2015} and \citet{chauvin2015}, but we summarize here the details that led to the definition of the final sample used in our work.

The target selection was originally based on a large compilation of $\sim$1000 young nearby stars that were selected in preparation of the VLT/SPHERE SHINE (SpHere INfrared survey for Exoplanets; Chauvin et al. in prep.) survey. The age determination was based on various youth indicators including lithium content, Ca~{\sc ii} H and K line chromospheric emission, H$\alpha$ emission, X-ray activity, rotation rate, and kinematics. From this initial compilation, we selected solar-type stars (FGK; 0.4 mag $\leq B -V \leq$ 1.2 mag) within a distance horizon of 100~pc and with an age $\leq$~200~Myr, in order to be sensitive to planetary-mass objects in the 50--1000~AU range. An additional cutoff at $R \leq 9.5$ was applied, corresponding to the limit for full-performance of the visible wavefront sensor of VLT/SPHERE expected at that time \citep{fusco2006}. Known spectroscopic and visual binaries ($\leq$~6\as) were also removed to have a sample as free as possible from dynamical perturbations to planets in wide orbits (see Sect.~\ref{sec:planet_population_synthesis_models}), as well as to avoid any decrease in performance of the adaptive optics system related to the presence of a stellar companion. Although spectroscopic binaries do not pose technical difficulties for wavefront sensing, they were discarded to focus the study on single stars and stars in extremely wide binary systems. The frequency of circumbinary giant planets has recently been the subject of a dedicated study \citep{bonavita2016}.

The list was then cross-correlated to the targets observed in previous high-contrast imaging surveys with sensitivities similar to the expected NaCo-LP sensitivity \citep{masciadri2005,lowrance2005,kasper2007,biller2007,lafreniere2007,chauvin2010}, resulting in a selection of 110 targets observable from the VLT and that were never previously observed at high-contrast. From this final selection, 85 were observed with VLT/NaCo from 2009 to 2013, while the other 24 could not be observed due to weather or technical losses. The results of the observations and a first statistical analysis of the NaCo-LP observed sample are presented in \citet{chauvin2015} and \citet{reggiani2016}.

For the final analysis that we present in this work, the observed sample was combined with the results from direct imaging surveys published prior to the start of the NaCo-LP, as well as surveys subsequently published \citep{heinze2010,vigan2012,rameau2013,biller2013,brandt2014}. We also added the results of an unpublished HST/NICMOS survey\footnote{HST, cycle 13, PI Inseok Song, \href{https://archive.stsci.edu/proposal_search.php?id=10176&mission=hst}{GO10176}: \emph{``Coronagraphic Survey for Giant Planets Around Nearby Young Stars''.}} by Song and collaborators and for which we performed a dedicated analysis. The candidates' identification and follow-up of this program was performed with VLT/NaCo (\citealt{chauvin2010}; Song, Farihi priv. com.). Only surveys targeting solar-type stars were included, and we did not include surveys targeting only stars with debris disks \citep{apai2008,wahhaj2013a} to avoid any significant astrophysical bias. However this selection criterion does not mean that we do not have debris disk targets in our sample. We simply included only surveys that did not use the presence of a known debris disk as their primary selection criterion. The color, $R$-mag, and distance criteria were also applied to the archival data, but we did not apply the age criterion so as to increase the number of targets in the sample. Table~\ref{tab:surveys} presents a summary of these surveys including the telescope and instrument, the band of the observations, the number of targets in their respective sample ($N_{\rm survey}$), and the number that we use in the current work ($N_{\rm NaCo-LP}$). From here on, we refer to the combination of the observed and archive samples as the full NaCo-LP sample, which comprises a total of 199 stars, some of which were observed in several surveys.

Ages and other stellar parameters for the literature surveys were rederived in an homogeneous way following the procedures described in \citet{desidera2015} for the NaCo sample. However, the ages of some young moving groups were updated, following the recent results by \citet{bell2015} and correspondingly the ages of stars based on indirect methods (e.g., lithium) taking as reference moving groups members were also updated. Further details on the adopted ages for individual groups are described in \citet{bonavita2016}. The ages of the targets from \citet{desidera2015} were also revised following the above procedure. The properties of the sample are summarized in Fig.~\ref{fig:sample_histograms}, and all the values are detailed in Appendix~\ref{sec:full_sample}.

From Fig.~\ref{fig:sample_histograms} we notice that the metallicity of the sample is quite narrowly distributed around solar value. This is the result of the fact that young stars in solar vicinity have typically metallicity close to the solar values \citep[see, e.g.,][]{santos2008,dorazi2009,biazzo2012}. Furthermore, the few stars with metallicity significantly above or below solar are typically among the oldest in the sample, and therefore carry limited information in the statistical analysis. This metallicity distribution and in particular the lack of super-metal-rich young stars should be taken into account in the interpretation of the statistical results of imaging surveys and in the comparison with the samples observed with other techniques like radial velocity, characterized by a broader metallicity distribution.

\section{Detection limits}
\label{sec:detection_limits}

\begin{figure}
    \centering
    \includegraphics[width=0.5\textwidth]{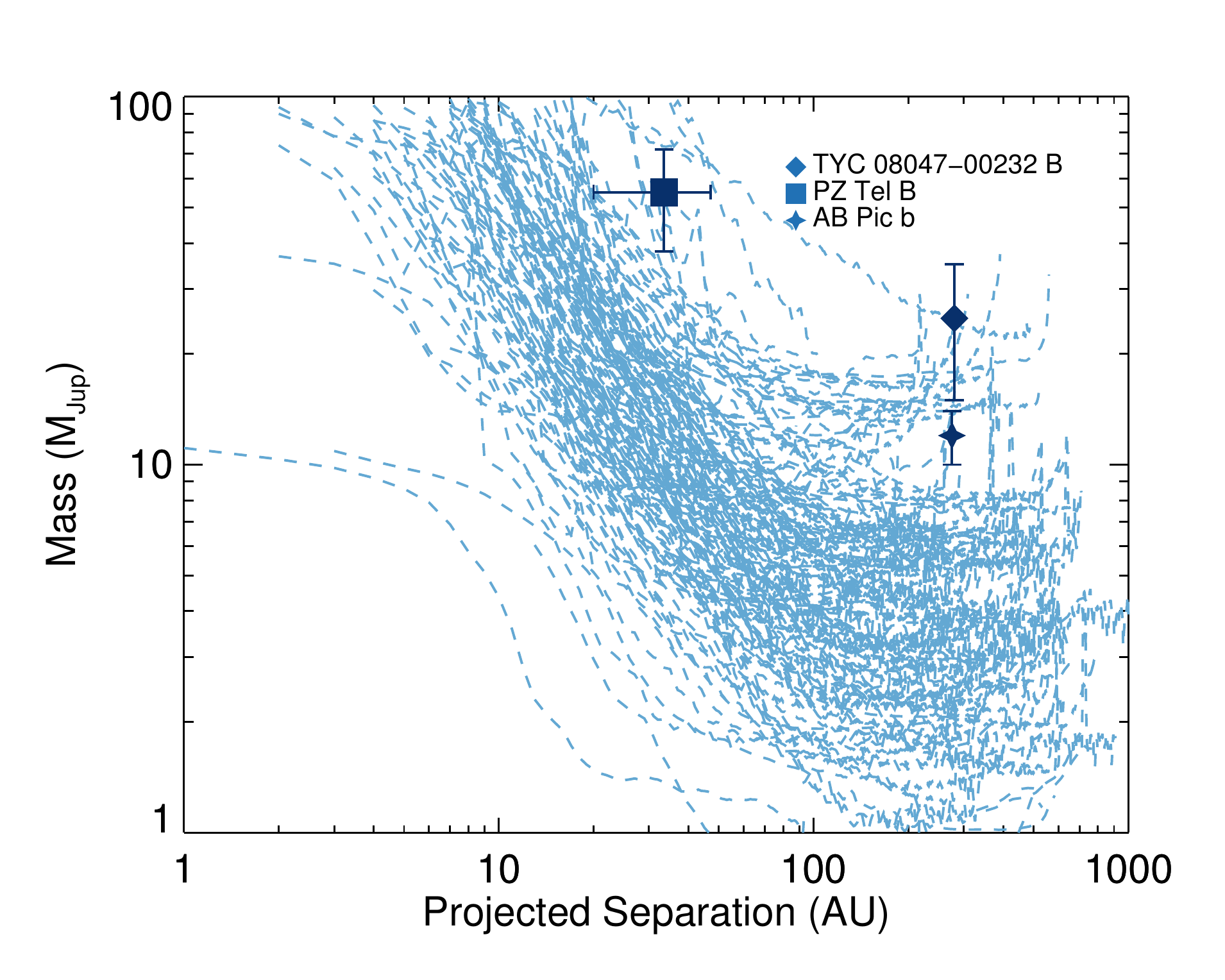}    
    \caption{Summary plot of all the mass detection limits considered in this work (blue dashed lines). The big dot points correspond to the detections of sub-stellar companions that have been considered in the analysis (see Sect.~\ref{sec:substellar_companions}). With an estimated mass of $0.1095 \pm 0.0022$~\MSun, HD~130948~BC lies outside the considered mass range, and it is therefore not shown.}
    \label{fig:1dlim}
\end{figure}

Detection limits for the various surveys were collected from their respective authors. The challenges of defining a reliable detection limit in high-contrast imaging have already been underlined by several authors in the case of angular differential imaging \citep[][]{marois2008b,mawet2014} or its combination with spectral differential imaging \citep{rameau2015,vigan2015}. To simplify the combination of many different detection limits from various surveys in our work, we adopt the commonly used 5$\sigma$ threshold for all detection limits. Although this may not be completely accurate, there is no simple alternative when considering many surveys, unless one has the capability of reprocessing all the corresponding data in a consistent manner. The only exception to this 5$\sigma$ threshold is the \citet{biller2013} survey, where the detection limit is given by a 95\% completeness curve, which represents the delta magnitude limit above which 95\% of all objects in the field should be detected at a given separation. \citet{wahhaj2013b}, who defined this metric, showed that in most cases it agrees well (at the $\sim$10\%--30\% level) with the nominal 5$\sigma$ contrast curves. 

While collecting all these detection limits, we realized that there is currently no public database that gathers all the results of past direct imaging surveys. In particular, the position and status of all companion candidates in the field is not always reported or easily accessible in machine-readable format. The availability of such data is of prime importance for the large imaging surveys being performed with VLT/SPHERE and Gemini/GPI, but also for the surveys that will be performed in the future. To remedy this lack, we associate the current paper with the release of DIVA (\url{http://cesam.lam.fr/diva/}), a public database designed to contain results from published direct imaging surveys: images, detection limits, signal-to-noise ratio maps, position, and status (when available) of detected companion candidates. To simplify and standardize these heterogeneous data, all the relevant information is packaged into HCI-FITS files, a standard for high-contrast imaging data recently proposed by \citet{choquet2014a} in the context of the ALICE\footnote{\url{https://archive.stsci.edu/prepds/alice/}.} (Archival Legacy Investigations of Circumstellar Environments) project \citep{choquet2014b,soummer2014}.

All the 5$\sigma$ detection limits and detection maps calibrated in contrast were then converted into companion mass using the AMES-COND2003 evolutionary models \citep{baraffe2003} specifically calculated in the filters of the observations (Fig.~\ref{fig:1dlim}). The post-formation luminosity of giant planets is likely to be the result of the way the material is accreted onto the protoplanetary core,  resulting in a broad range of possible luminosities for a similar mass \citep{fortney2008,spiegel2012,mordasini2013}. However, comparisons of the luminosity of some giant planets with sets of evolutionary tracks or formation models \citep[e.g.,][]{janson2011,bonnefoy2013} tend to show that they do not appear to be compatible with very low initial entropy models (``cold-start'') but instead with intermediate or high initial entropy models (``warm-start'' or ``hot-start''). For the mass conversion we therefore use the AMES-COND2003 models which correspond to the highest possible initial entropy, but keeping in mind that it is the most favorable scenario. Future studies will need to address the initial entropy problem in large direct imaging surveys more specifically.

\section{Sub-stellar companion detections}
\label{sec:substellar_companions}

During the course of the NaCo-LP, we detected twelve previously unknown binaries \citep{chauvin2015} and one white dwarf \citep{zurlo2013}, but no new sub-stellar companions. However, our full statistical sample includes four known companions discovered by other surveys and that must be considered in the analysis. We summarize below the main properties of these companions:

\begin{itemize}
    \item \object{GSC~08047-00232}~B (\object{TYC~8047-0232}~B) is a $25 \pm 10$~\MJup BD with a derived spectral type of M9.5$\pm$1 \citep{chauvin2005a} at a projected separation of $\sim$280~AU. It is a probable member of the Tucana-Horologium association, with an age of 10--50~Myr. While the age indicators of GSC~08047-00232 are consistent with other K-type members of Tuc-Hor, its available kinematics are not a strong match to the bulk of the association. For the purpose of this paper we keep this star as a member of Tucana-Horologium;
    \item \object{AB~Pic}~b (\object{HIP~30034}~b) is a 10--14~\MJup object at a projected separation of 275~AU \citep{chauvin2005b,bonnefoy2010}. The object AB~Pic b could alternatively be a member of the similarly aged, but less well defined, Carina association \citep{malo2013}. This has no impact because the ages of Tuc-Hor and Carina are the same within uncertainties;
    \item \object{PZ~Tel}~B (\object{HIP~92680}~B) is a $50^{+13}_{-8}$~\MJup BD companion \citep{biller2010, maire2016} orbiting on a very eccentric orbit (e>0.66) at 49--77~AU. It is a member of the $\beta$~Pictoris moving group.
    \item \object{HD~130948}~BC (\object{HIP~72567}~BC) is a BD binary discovered by \citet{potter2002}, with a total mass estimated at $0.1095 \pm 0.0022$~\MSun \citep{dupuy2011} and orbiting at a projected separation of $\sim$50~AU from the Sun-like star HD~130948~A. As this is a hierarchical triple system, it is considered in our sample as a binary with masses $M_A$ and $M_{BC}$. As the secondary has a mass outside the mass ranges of interest of this study, we will not consider this as a detection in our analysis.
\end{itemize}

\section{Planet population synthesis models}
\label{sec:planet_population_synthesis_models}

\begin{figure*}
    \centering
    \includegraphics[width=0.49\textwidth]{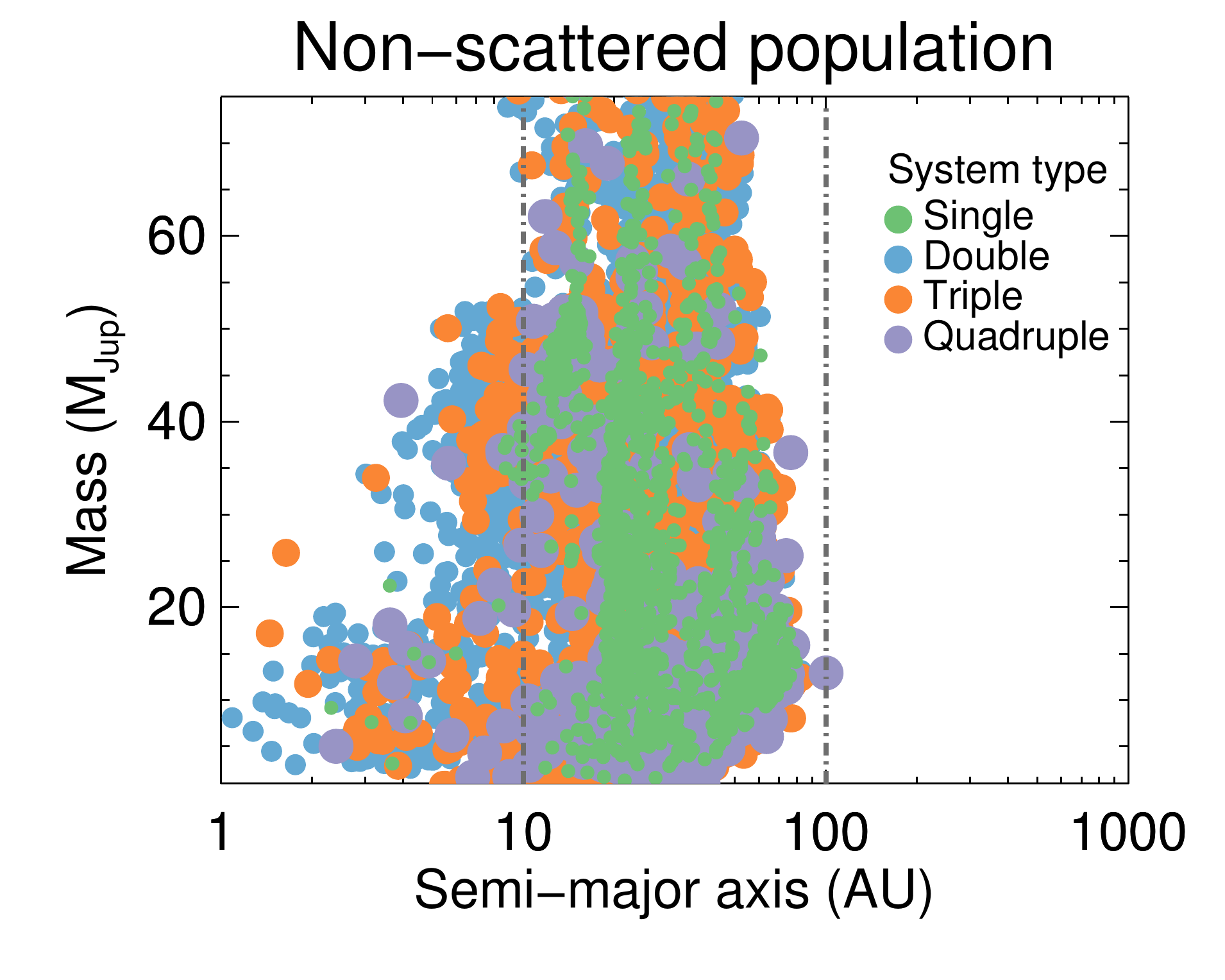}    
    \includegraphics[width=0.49\textwidth]{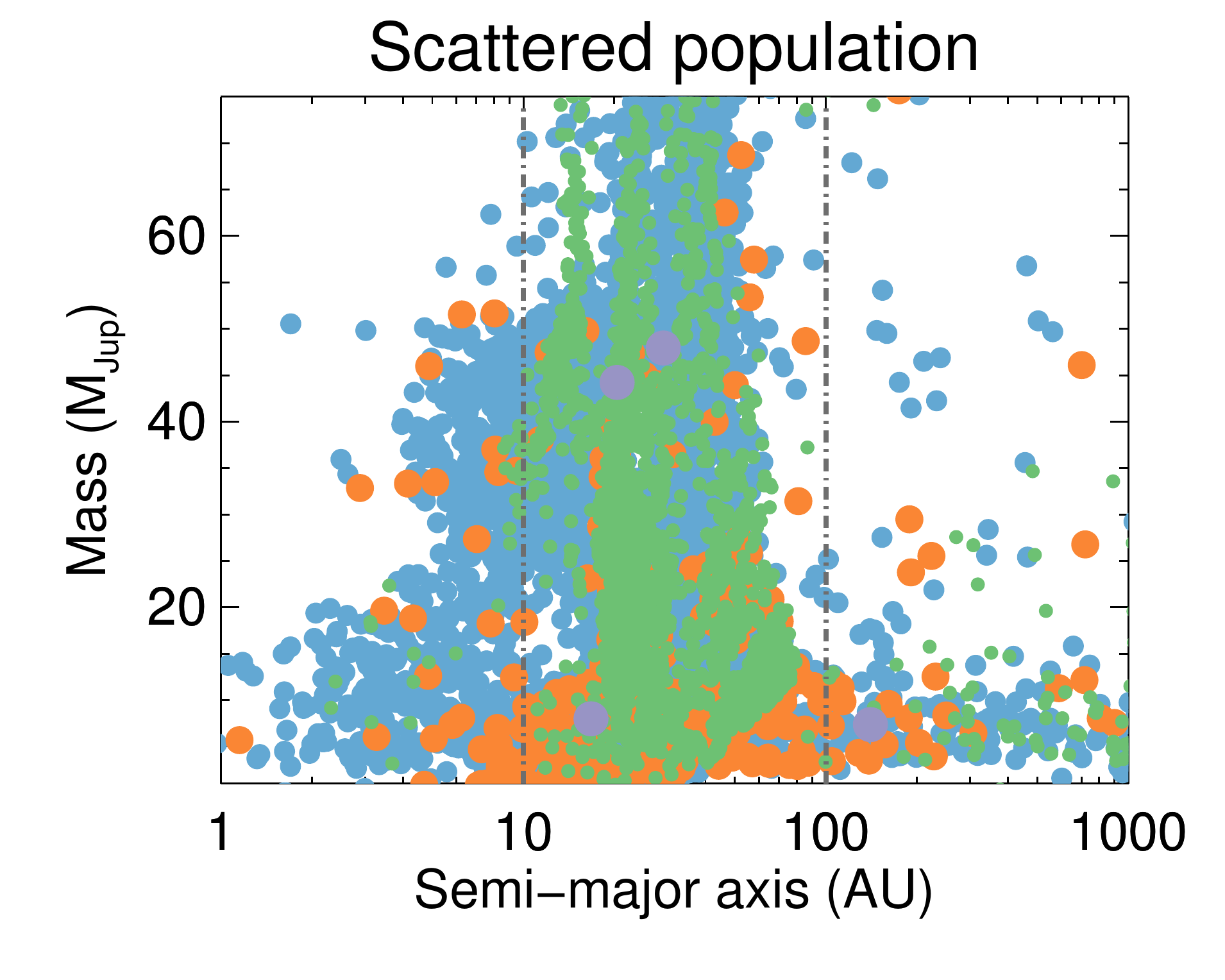}
    \caption{Mass (\MJup) versus physical separation (AU) for all the companions in the synthetic planet populations generated as described in Sec.~\ref{sec:planet_population_synthesis_models}. Left panel: non-scattered population \citep{forgan2013}. Right panel: same population including scattering \citep{forgan2015}. Different colors are used for single (green), double (blue), triple (orange), and quadruple (purple) systems. The size of the symbols increases from single to quadruple systems to make the plots easier to read.}
    \label{fig:synth_pop}
\end{figure*}

The standard planet formation scenario is typically referred to as the core accretion model \citep{safronov1972}. This involves the growth of micron-sized dust grains to form kilometre-sized planetesimals that then collide and grow to form planetary-mass bodies \citep{wetherill1990,kokubo1998}. If these become sufficiently massive prior to the dispersal of the disk gas, they may accrete a gaseous envelope \citep{mizuno1980} to become a gas giant planet \citep{Pollack1996}. If not, they remain as rocky and icy planets with masses similar to that of the Earth \citep{chambers1998}.

The detection of giant planets on wide orbits, however, provides a challenge to the standard core accretion scenario, as the growth time for cores at these radii should far exceed the gas disk lifetime \citep{levison2001}. The accretion of centimeter-sized pebbles has been proposed as a possible way of rapidly forming gas giant cores \citep{lambrechts2012}, but this model is relatively young so further exploration of this mechanism is still necessary. An alternative scenario involves direct gravitational collapse in massive, self-gravitating protostellar disks \citep{kuiper1951,boss1998}. The basic idea is that if a protostellar disk is both massive and cold, it becomes susceptible to the growth of gravitational instability \citep{toomre64}. If these disks can cool rapidly \citep{gammie01,rice03}, or maintain sufficiently high accretion rates while remaining quasi-isothermal \citep{Kratter2011}, then this can lead to fragmentation to form bound objects that may then collapse to form planetary-mass bodies, brown dwarfs, or low-mass stars \citep{stamatellos2009a}.

Furthermore, it has been suggested \citep{nayakshin2010} that within these fragments, dust grains may grow and sediment to form cores, and that the fragments may rapidly migrate into the inner disk. If the contraction of the fragment is slow enough, and the migration and sedimentation fast enough, then the fragment may be stripped of gas by tidal interactions with the host star and this process could in principle form both close-in, and wide orbit, planets \citep{nayakshin2016b}.   

It is therefore possible to generate a synthetic population of planets -- or brown dwarfs -- that form via disk fragmentation, undergo grain sedimentation, and evolve via migration in the protostellar disks and, potentially, via tidal interactions with the host star. We briefly describe the main parameters of the models below, but the reader is encouraged to read \citet{forgan2013} and \citet{forgan2015} for a complete description. The models combine:

\begin{enumerate}
    \item semi-analytic self-gravitating disk models with photoevaporation \citep{rice2009,owen2011},
    \item analytic fragmentation criteria which measure the local Jeans mass inside a spiral perturbation \citep{forgan2011}, 
    \item a system of fragment evolution equations which follow the evolution of the fragment's dust and gas components \citep{2010MNRAS.408L..36N,2010MNRAS.408.2381N,2011MNRAS.413.1462N},
    \item equations that describe the orbital evolution of the protoplanet via disk migration \citep{kley2012,baruteau2013}, and
    \item equations describing the tidal disruption of the planetary embryo if it exceeds its Hill radius \citep{faber2005}.
\end{enumerate}

\begin{figure}
    \centering
    \includegraphics[width=0.49\textwidth]{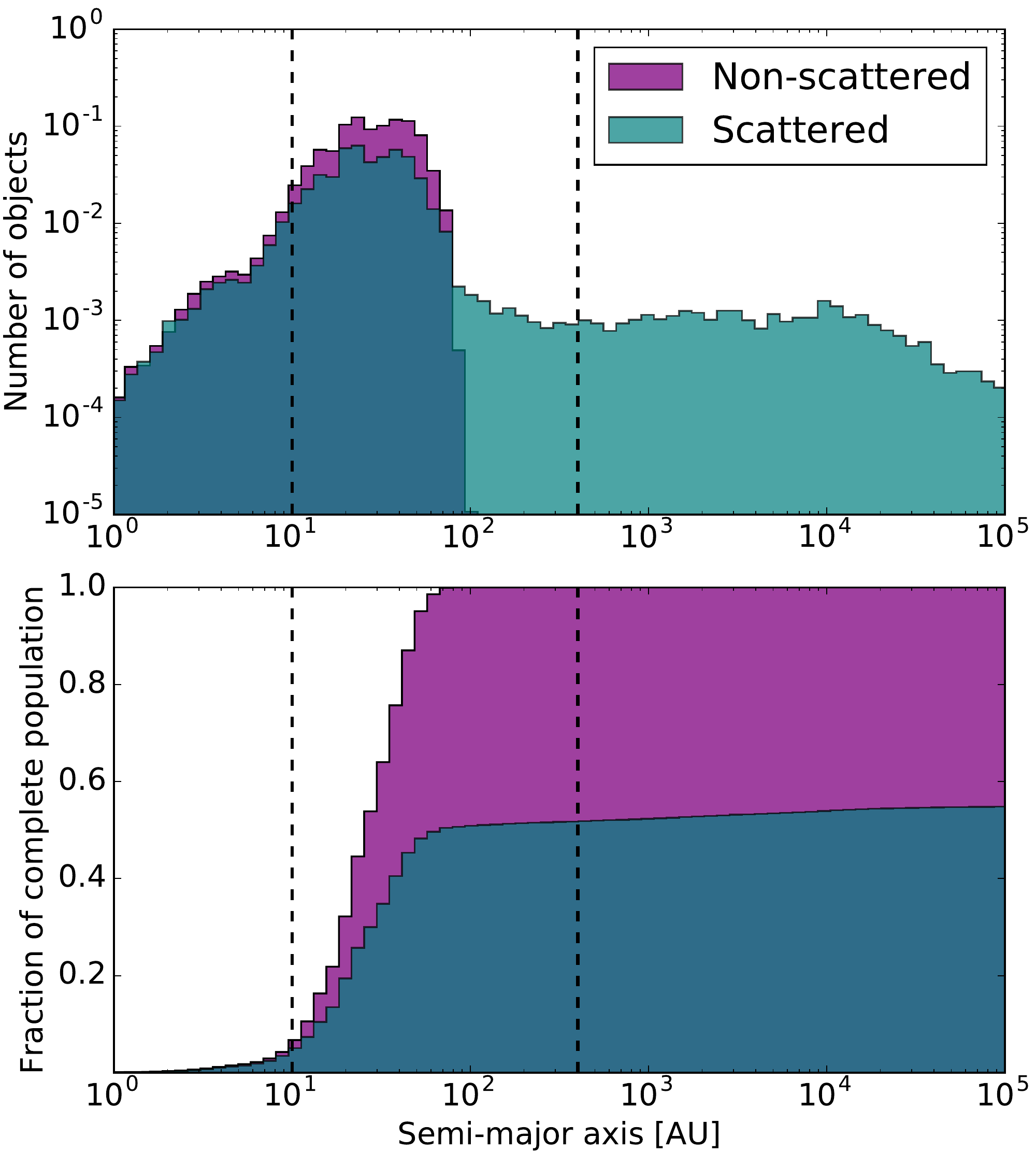}
    \caption{Histogram (top) and cumulative histogram (bottom) of the population before and after scattering as a function of semi-major axis, normalized with respect to the total number of planets in the population. Only objects below 75~\MJup are considered. Approximately 50\% of the planets are ejected from the systems during the scattering, resulting in a much smaller total number of planets in the scattered population. The vertical dashed lines at 10 and 400~AU mark the semi-major axis range where the average detection probability of our observations is $\sim$50\% or more for massive companions.}
    \label{fig:sma_histo}
\end{figure}

\begin{figure}
    \centering
    \includegraphics[width=0.49\textwidth]{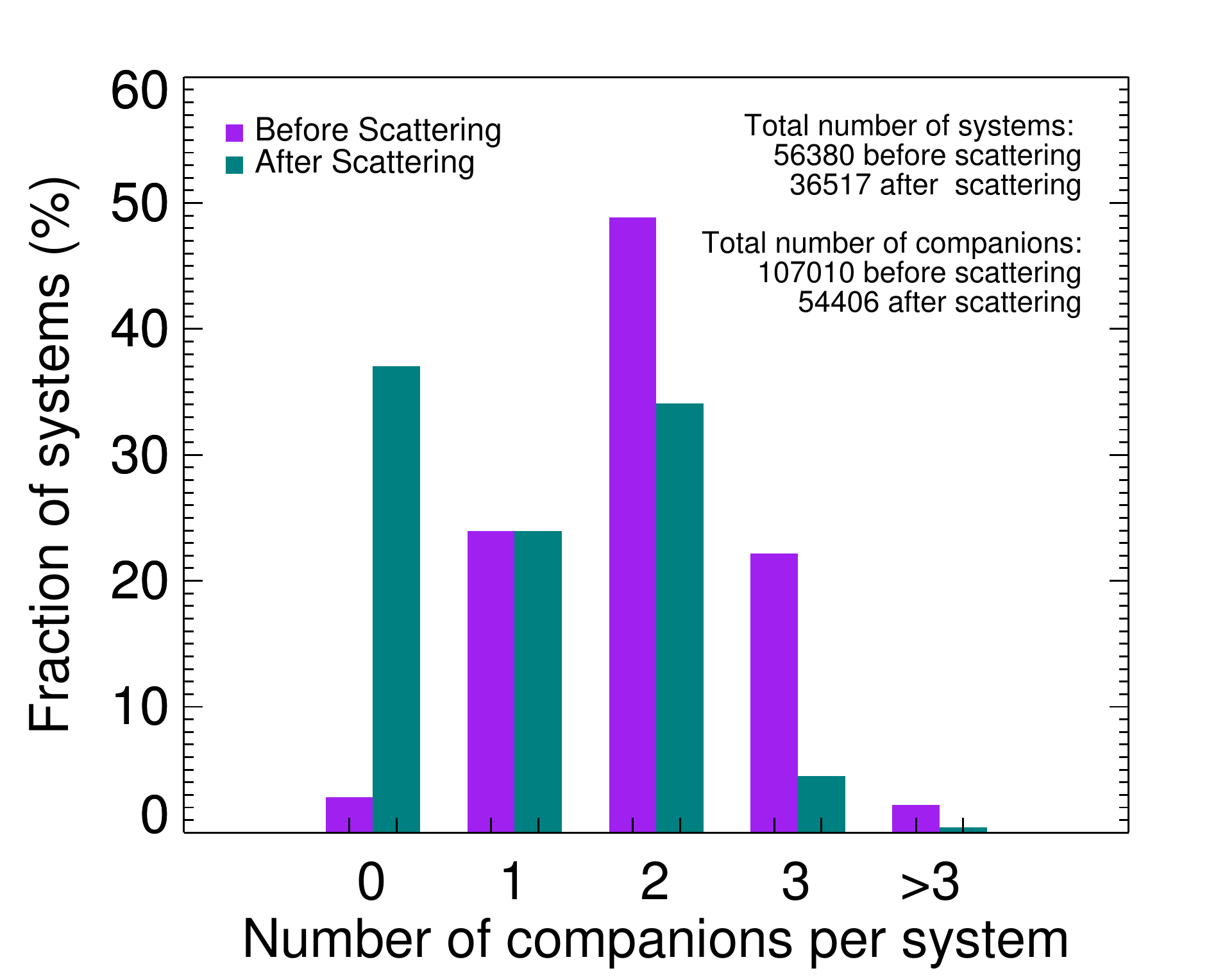}
    \caption{Fraction of systems in the scattered and non-scattered populations as a function of the number of planets in the system. The total number of systems and planets before and after scattering is displayed in the top-right corner of the plot. The bins are normalized to the total number of systems in each of the two populations independently. The $\sim$2\% of systems with zero planets before scattering correspond only to cases where all the companions in the systems have masses above $75~M_{Jup}$, while for the scattered case it also includes systems where all the companions were ejected as a result of the scattering process.}
    \label{fig:sys_barplot}
\end{figure}

Once fragmentation has occurred, the embryos continue to collapse in a manner analogous to protostellar collapse, while their internal dust population grows and begins to sediment towards the pressure maximum at the center. This sedimentation is suppressed partially by gas turbulence and high-velocity collisional destruction, and eventually by evaporation as the gas temperature continues to increase. The embryo migrates radially inward according to the local disk properties, and can be tidally disrupted as it overflows its own Roche lobe. Indeed, almost half of all embryos are destroyed during this process.

In this work, we assume a disk with maximum radius of 100 au, with a surface density profile $\Sigma \propto r^{-1}$, and with solar metallicity.  The stellar mass and disk mass are varied for each iteration of the model, as is the X-ray luminosity of the star, which affects the timescale for disk photoevaporation and hence sets a time limit on the fragments' evolution.  The resulting population generated is described in detail in section 3.1 of \citet{forgan2013}.

The population synthesis model yields (mainly) relatively massive gas giants and brown dwarfs, with the majority of these objects residing at semi-major axes above 30~AU. As this model does not consider the dynamical evolution of these bodies after the protostellar disk has dispersed, the systems for which more than two embryos have formed are evolved for $10^6$ years using n-body simulations to ascertain the effect of object-object scattering on the planetary orbital parameters \citep{forgan2015}. Although each system is evolved for a time that is smaller than the observed ages of objects which we would like to compare to disk fragment models, this relatively low simulation time was used partly to reduce computational expense and partly because systems that produce scattering events express this instability within a few tens of thousands of years \citep{chambers1996,chatterjee2008}. As such, the results should reasonably reflect the likely outcome of fragment-fragment interactions. Having said this, we must note that secular evolution of these systems may occur over timescales greater than $10^6$ yr \citep[e.g.,][]{veras2009}.  As a result, we may underestimate the number of bodies ejected from their host star systems, and we may also underestimate the final semimajor axis of the bodies that do remain bound.

This procedure therefore gives two data sets to consider: the first is the output of the population synthesis model \citep[hereafter non-scattered population;][]{forgan2013}, and the second is the dynamically evolved, post-disk dispersal population \citep[hereafter scattered population;][]{forgan2015}. 

Figure~\ref{fig:synth_pop} shows the output of the population synthesis models for both the scattered (left panel) and non-scattered (right panel) population. We can see that for both before and after scattering, the majority of the population resides at semi-major axes >20~AU, with masses of the order of one Jupiter mass ($\MJup$) and larger. This is despite the model's ability to tidally downsize the bodies into the Neptune and Earth mass regime, which some authors predict \citep{nayakshin2010}, but we find such events are statistically rare enough not to occur, despite synthesising a total of one million disk fragments. Crucially, whatever subsequent modification occurs to the bodies after fragmentation, the process preferentially generates large populations of bodies with masses above 1~\MJup and semi-major axes above 20~AU, well within the sweet spot of direct imaging observation sensitivity reported in the present work. 

Figures~\ref{fig:sma_histo} and \ref{fig:sys_barplot} show the evolution of the two populations in terms of semi-major axis distribution and repartition of the number of planets in the systems. For the semi-major axis, the scattering creates a tail of planets that go much beyond 100~AU, but the population remains largely dominated by a large peak in the 10--100~AU range. However, the cumulative histogram shows that $\sim$50\% of the planets are ejected from the systems during the scattering. This process affects mainly systems with three or more planets (Fig.~\ref{fig:sys_barplot}), which are sometimes transformed into systems with one or two planets, but which more generally end up with not planets at all.

We should note of course that the population synthesis models of \citet{forgan2013} are the first steps towards a complete description of disk fragmentation and the tidal downsizing process. The post-processing of the model's output by \citet{forgan2015} highlights the importance of fragment-fragment interactions in setting the distribution of orbital parameters (especially eccentricity). However, these interactions only occur after the disk dissipates -- interactions between fragments while still embedded in the disk are neglected. Simulations of fragmenting disks demonstrate that these early fragment interactions can determine whether a fragment survives tidal downsizing to become a detectable sub-stellar object (\citealt{stamatellos2009a,stamatellos2009b}; Hall et al. submitted).

The model does not permit accretion of further gas from the disk by the fragment, and therefore the masses derived by the model are lower limits. Mass accretion can encourage the opening of gaps in the disk and reduce the migration speed, which allows more massive objects to remain at larger semi-major axis. That being said, \citet{Baruteau2011} and \citet{Zhu2012} show that migration rates in self-gravitating disks are rapid compared to theoretical calculations based on low-mass disks, and \citet{malik2015} show that this can frustrate gap opening even for high fragment masses. If fragments do survive the tidal downsizing process, these results suggest such objects are likely to be more massive, and hence easier to detect than the \citet{forgan2013} model currently predicts.

The accretion of solid material \citep[e.g.,][]{helled2006} is also not modeled, which in turn affects the microphysics of dust settling in the fragment \citep{nayakshin2016a}. As this is responsible for the formation of solid cores in the GI formalism, this affects the survival rates of disk fragments. The physics of fragment evolution is at its heart the study of dust-gas interactions in a collapsing environment, a process that is challenging to describe semi-analytically. The interested reader should consult \citet{nayakshin2016b} for a review of the multiple physical processes at play.

Finally, it is still somewhat unclear as to how stellar companions might influence the evolution of self-gravitating protostellar disks, which is one of the reasons why visual binaries with separations below 6\as have been removed from the sample. Much of the early work in this area investigated the role of stellar encounters, which indicate how the impulse induced by a companion can promote or inhibit fragmentation. \citet{Boffin1998} suggested that encounters might promote disk fragmentation, while more recent work has suggested that they will typically inhibit fragmentation \citep{Lodato2007,Forgan2009}. This latter work, however, considered disks that were initially sufficiently compact that fragmentation in isolation was unlikely. If disks can be sufficiently extended that they become irradiation dominated, then encounters may well trigger fragmentation in the outer regions of these disks \citep{Thies2010}. It is, however, still unclear as to whether or not such extended disks actually exist at such early times \citep{Maury2010}.

More recent studies now consider the secular evolution of self-gravitating protostellar disks by companions on closed orbits.  For example, \citet{Fu2017} demonstrated for isothermal disks that if the companion's orbit is sufficiently tilted compared to the disk plane, Kozai-Lidov oscillations can induce fragmentation even when the disk has a Toomre parameter $Q>2$.  Further work is required to fully ascertain the role of binary companions in disk fragmentation, with particular attention needing to be paid to disk thermodynamics and sweeping the relatively large parameter space of both binary and disk configurations.

\section{Statistical Analysis}
\label{sec:statistical_analysis}

\begin{figure}
    \centering
    \includegraphics[width=0.5\textwidth]{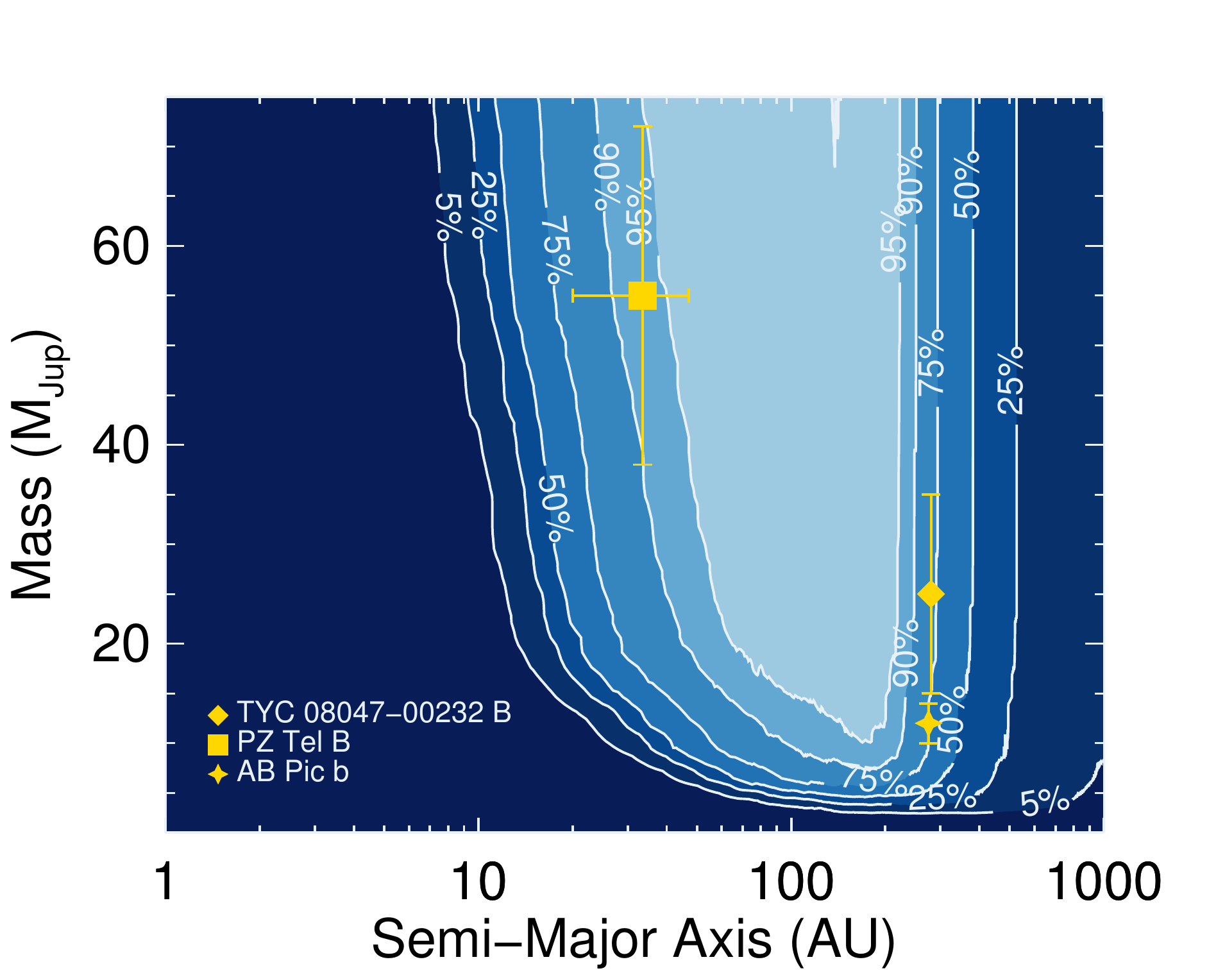}    
    \caption{Mean detection probability map for the complete sample of 199 stars as a function of mass and semi-major axis. In this plot we adopt the best estimate for the stellar ages. The four detections of the sample are over-plotted. The probability map assumes flat distributions of planets in mass and semi-major axis.}
    \label{fig:mean_prob}
\end{figure}

\begin{figure}
    \centering
    \includegraphics[width=0.49\textwidth]{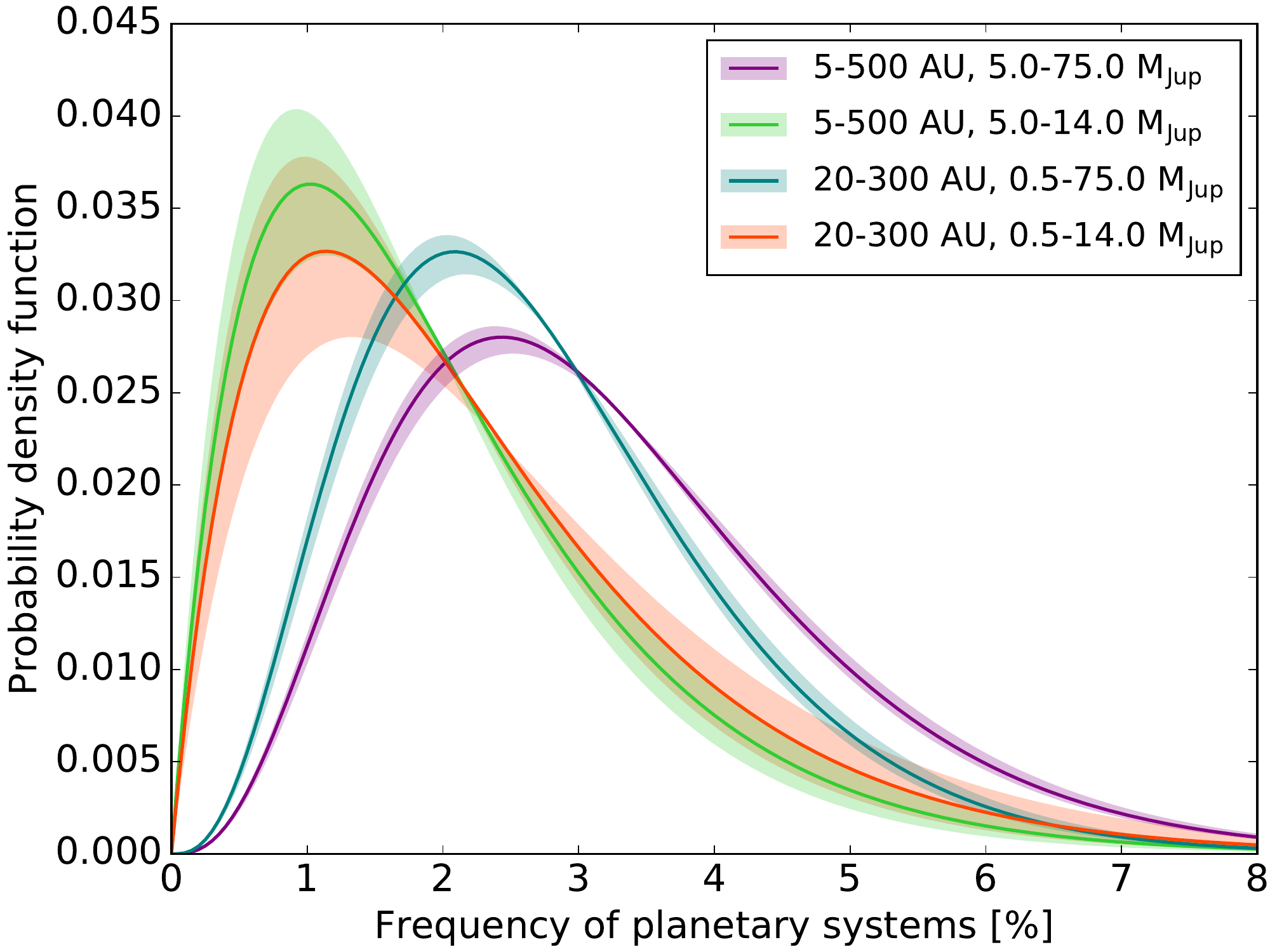}
    \caption{Probability distribution of the frequency of stars with at least one giant planet in the mass and semi-major axes ranges indicated in the top-right of the plot. The results are summarized in Table~\ref{tab:freq}. The plain lines represent the probability density function obtained when considering the best age estimate for all the stars in the sample, and the colored envelopes represent the variations when considering the minimum and maximum ages.}
    \label{fig:cal_frac}
\end{figure}

\begin{figure*}
    \centering 
    \includegraphics[width=0.48\textwidth]{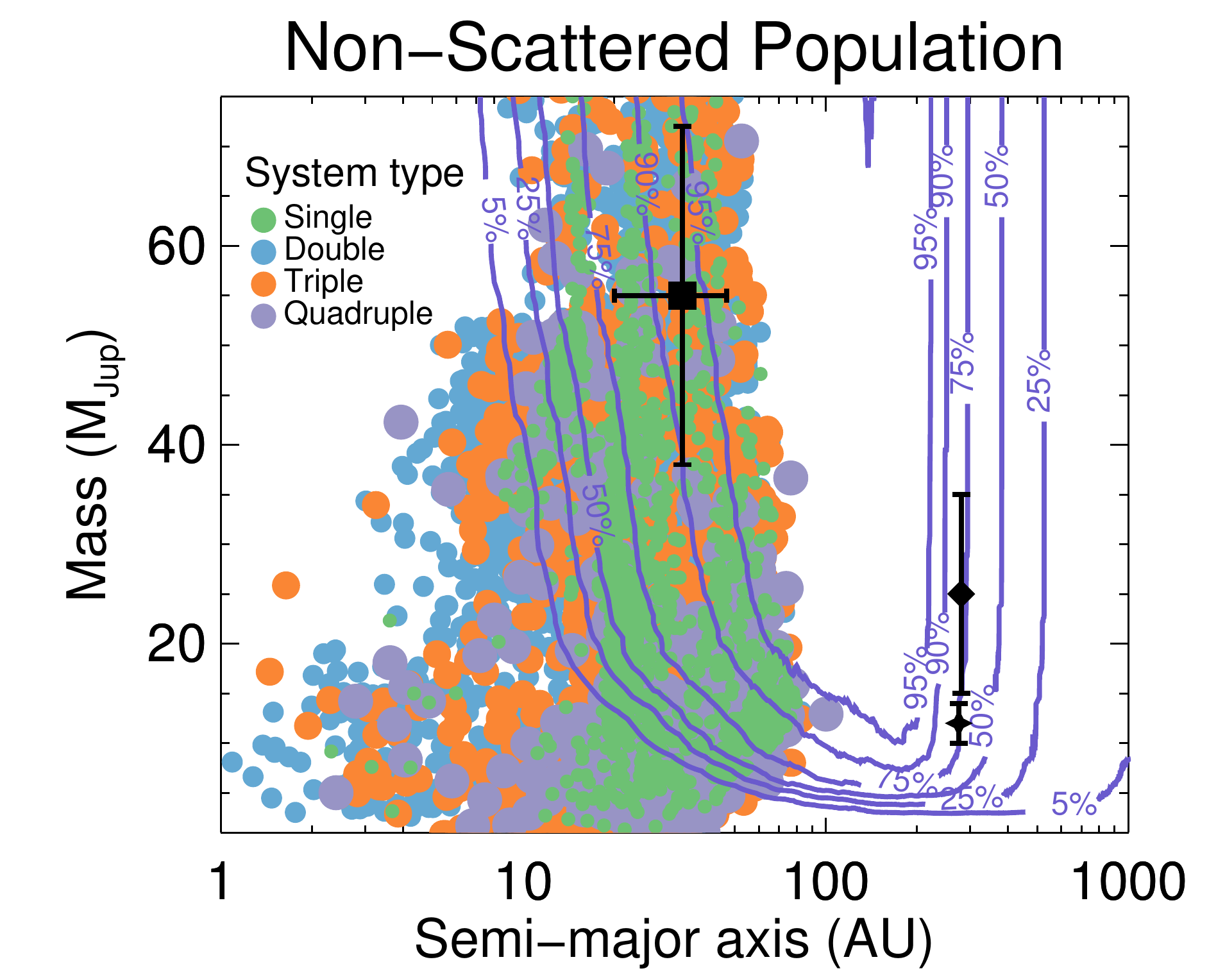}
    \includegraphics[width=0.48\textwidth]{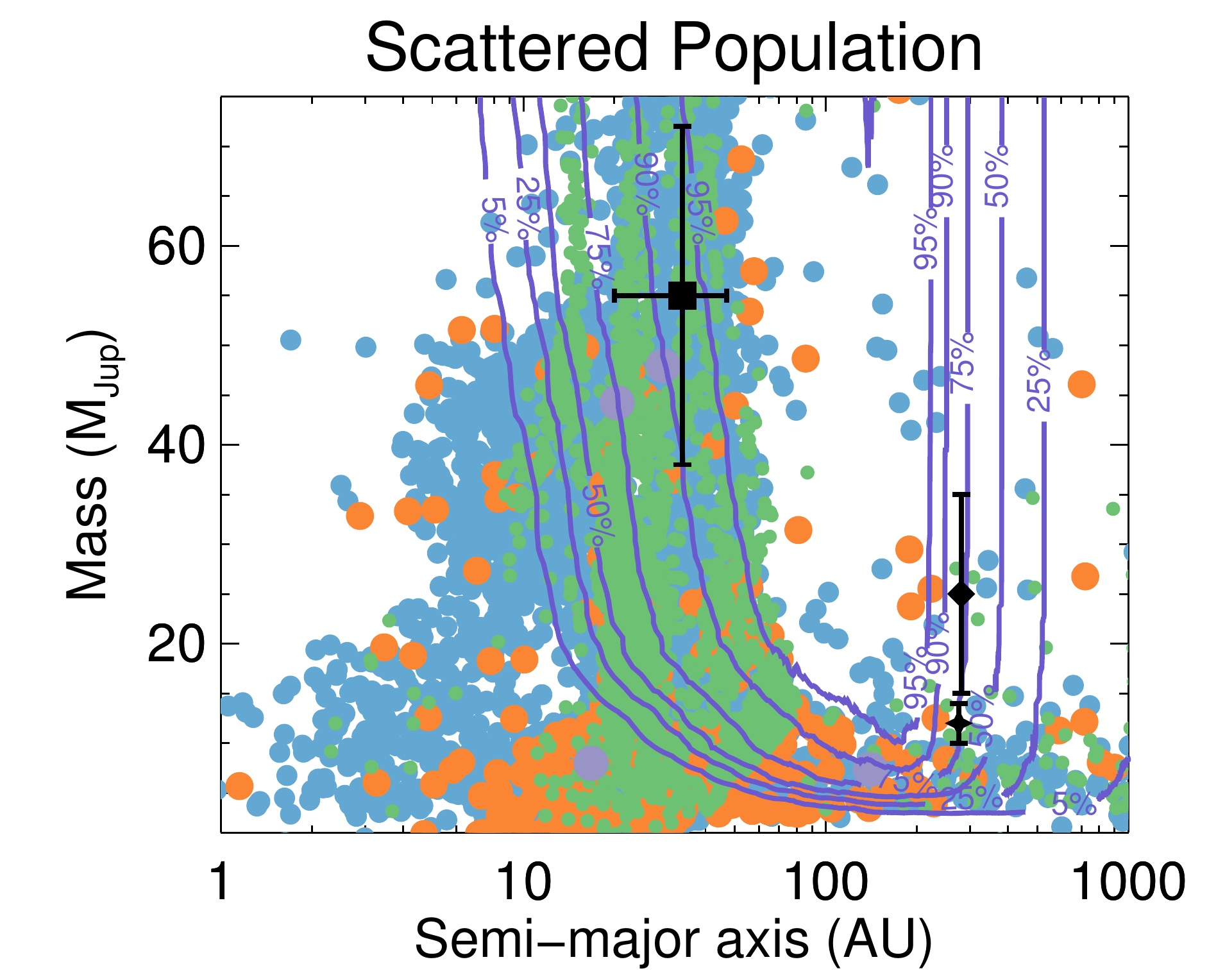}
    \caption{Comparison of the non-scattered (left) and scattered (right) GI populations \citep{forgan2013,forgan2015} with the mean detection probabilities of the observations (best age estimate). A significant fraction of the GI population could potentially be detected, even in the non-scattered case. The colors and symbols for the populations and the known companions are the same as in Fig.~\ref{fig:synth_pop} and \ref{fig:mean_prob}.}
    \label{fig:mod_cmp}
\end{figure*}

\subsection{Giant planet and brown dwarf frequency estimation} 
\label{sec:frequency_estimation}

\begin{table}
    \caption[]{Fraction of stars that host a planetary system, assuming linear flat mass and semi-major axis distributions.}
    \label{tab:freq}
    \centering
    \begin{tabular}{cccccc}
    \hline \hline
    SMA     & Mass               & $N_{det}$\tablefootmark{a} & $F_{best}$\tablefootmark{b} & \multicolumn{2}{c}{$[F_{min}, F_{max}]$\tablefootmark{c}} \\
    (AU)    & ($M_\mathrm{Jup}$) &                            & (\%)                        & CL=68\%      & CL=95\%       \\
    \hline
     5--500 &   5--75            & 3                          & 2.45                        & [1.70, 4.70] & [0.90, 6.95]  \\
     5--500 &   5--14            & 1                          & 1.00                        & [0.75, 3.30] & [0.25, 5.55]  \\
    20--300 & 0.5--75            & 3                          & 2.10                        & [1.50, 4.05] & [0.80, 5.95]  \\
    20--300 & 0.5--14            & 1                          & 1.15                        & [0.85, 3.65] & [0.30, 6.20]  \\
    \hline
    \end{tabular} 
    \tablefoot{Numbers are provided for the best stellar ages. Values for the minimum and maximum ages are provided in Appendix~\ref{sec:app_freq}. \tablefoottext{a}{Number of detections in the considered mass and semi-major axis (SMA) range.} \tablefoottext{b}{Best value of the planet frequency compatible with the observations.} \tablefoottext{c}{Minimum and maximum values of the frequency compatible with the results, for a given confidence level (CL).}}
\end{table}

The statistical analysis of the survey results has been done using the MESS code \citep[Multi-purpose Exoplanet Simulation System, see][]{bonavita2012} and its QMESS evolution \citep{bonavita2013}. The code allows a high level of flexibility in terms of possible assumptions on the synthetic planet population to be used for the determination of the detection probability. 
For consistency with previous work \citep[see, e.g.,][]{lafreniere2007,vigan2012,rameau2013} and to allow a straightforward comparison of the results, we started our analysis by generating a population of planets using a given set of assumptions on the planet parameter distributions, and comparing the detectability of this population of planets with our detection limits to obtain a mean probability of detection map. We assumed flat distributions in linear space for the mass and semi-major axis respectively, in the 1--75~\MJup range by steps of 1~\MJup and in the 1--1000~AU range by steps of 1~AU. Following \citet{hogg2010}, we adopt a Gaussian eccentricity distribution with $\mu =0$ and $\sigma = 0.3$. The implications of choosing this distribution over a flat eccentricity distribution are discussed in \citet{bonavita2013}. The mean sensitivity map that we obtain, plotted in Fig.~\ref{fig:mean_prob} for the optimal stellar ages of the sample, illustrates the high sensitivity of the full NaCo-LP sample of brown dwarfs and massive planets down to $\sim$10~\MJup in the 50--300~AU range. 

Following \citet{lafreniere2007} and \citet{vigan2012}, we can estimate the fraction of stars that host at least one brown dwarf or giant planet, F$_{ps}$, using the detection probability maps of individual targets and the confirmed detections in our sample (see Sect.~\ref{sec:substellar_companions}). Since we have no a priori information on F$_{ps}$, we assume a flat prior. The probability density function (PDF) of F$_{ps}$ in different semi-major axis and mass intervals is plotted in Fig.~\ref{fig:cal_frac}, and the results are summarized for the best stellar ages in Table~\ref{tab:freq} (results for the extreme ages are provided in Appendix~\ref{sec:app_freq}). As mentioned in Sec.~\ref{sec:substellar_companions}, HD~130948~BC is considered in our sample as a binary and is therefore not used in this analysis. The frequency values are in agreement with the results presented by \citet{galicher2016} for a sample of a  similar size to ours. They support a frequency between 1-2\% and 3-4\% for planetary-mass and brown dwarfs around FGK stars within 100~pc.

To remain as unbiased as possible when deriving the PDF of F$_{ps}$, this analysis assumes flat distributions in mass and semi-major axis (i.e., linear-flat priors). A more realistic approach would be to consider log-uniform priors, which would favor less massive, closer-in planets compared to distant giant planets. The main qualitative effect on our results would be to increase the number of planets that we cannot detect, and would therefore broaden significantly the PDFs showed in Fig.~\ref{fig:cal_frac}. Instead of relying on these modified priors, which are more realistic but still disconnected from the physics of planetary formation, we propose to use population synthesis models to constrain the planetary system frequency.

\subsection{Analysis based on population synthesis models}
\label{sec:synth_pop_mess}

\begin{figure*}
    \centering
    \includegraphics[width=0.48\textwidth]{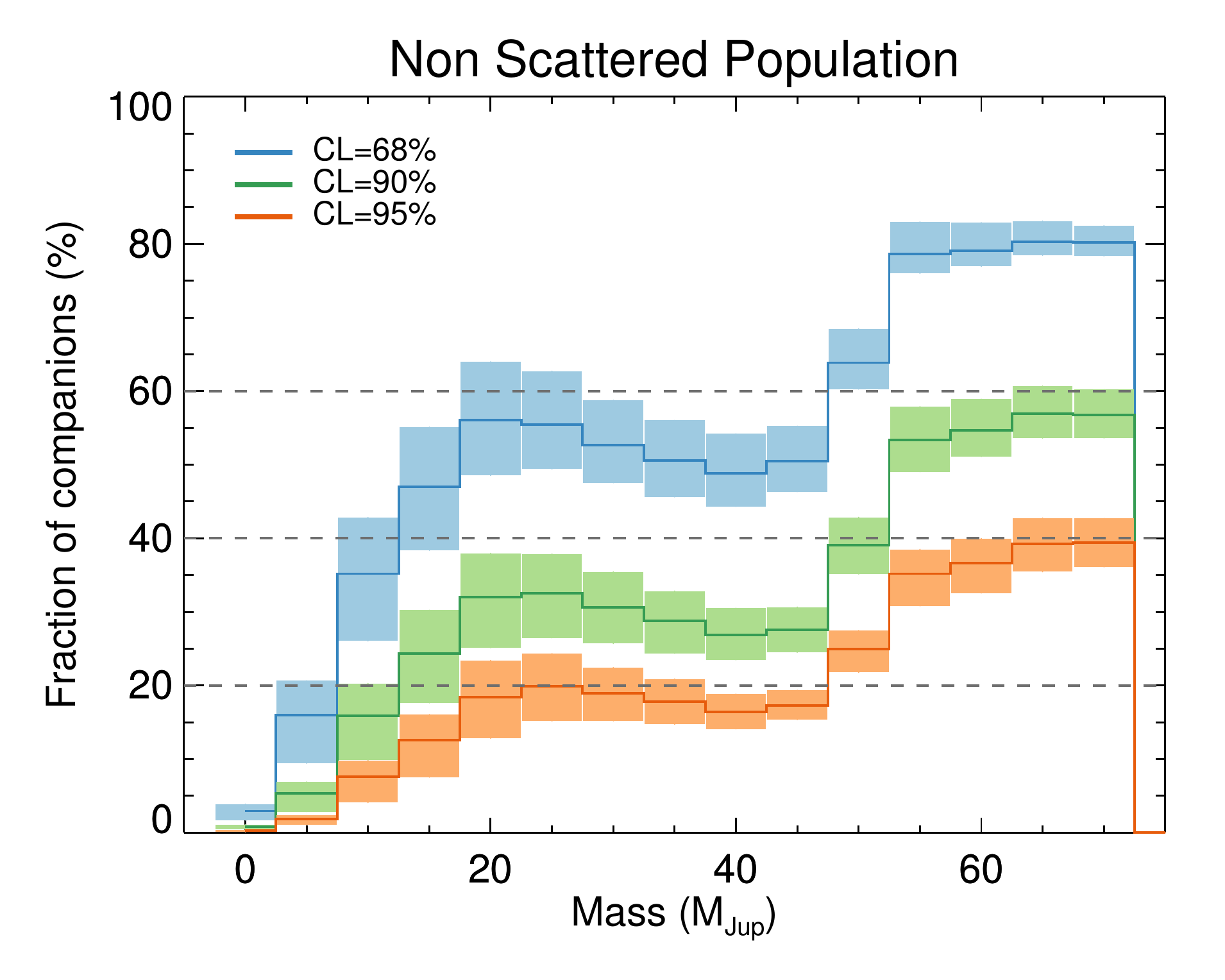}
    \includegraphics[width=0.48\textwidth]{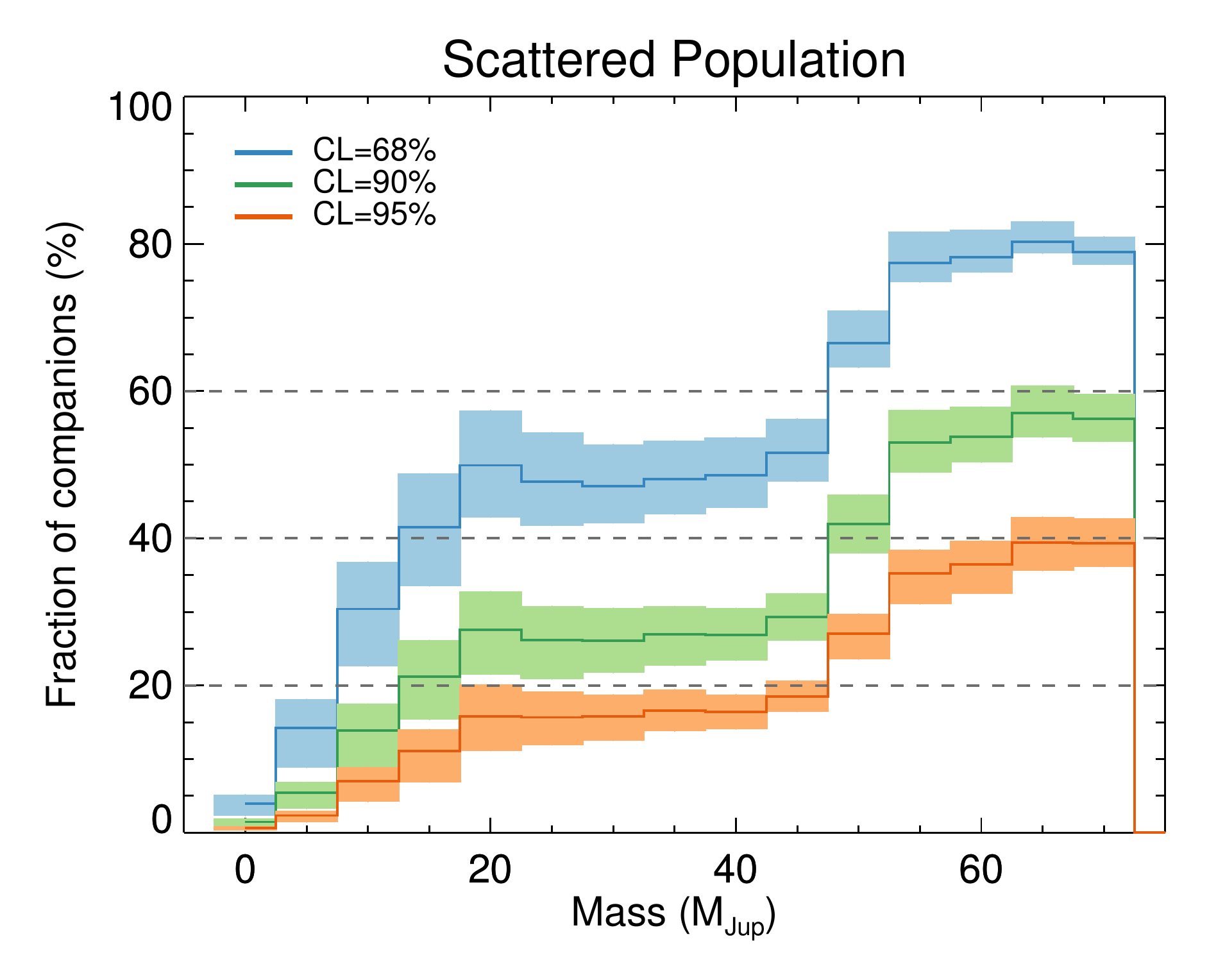} \\
    \includegraphics[width=0.48\textwidth]{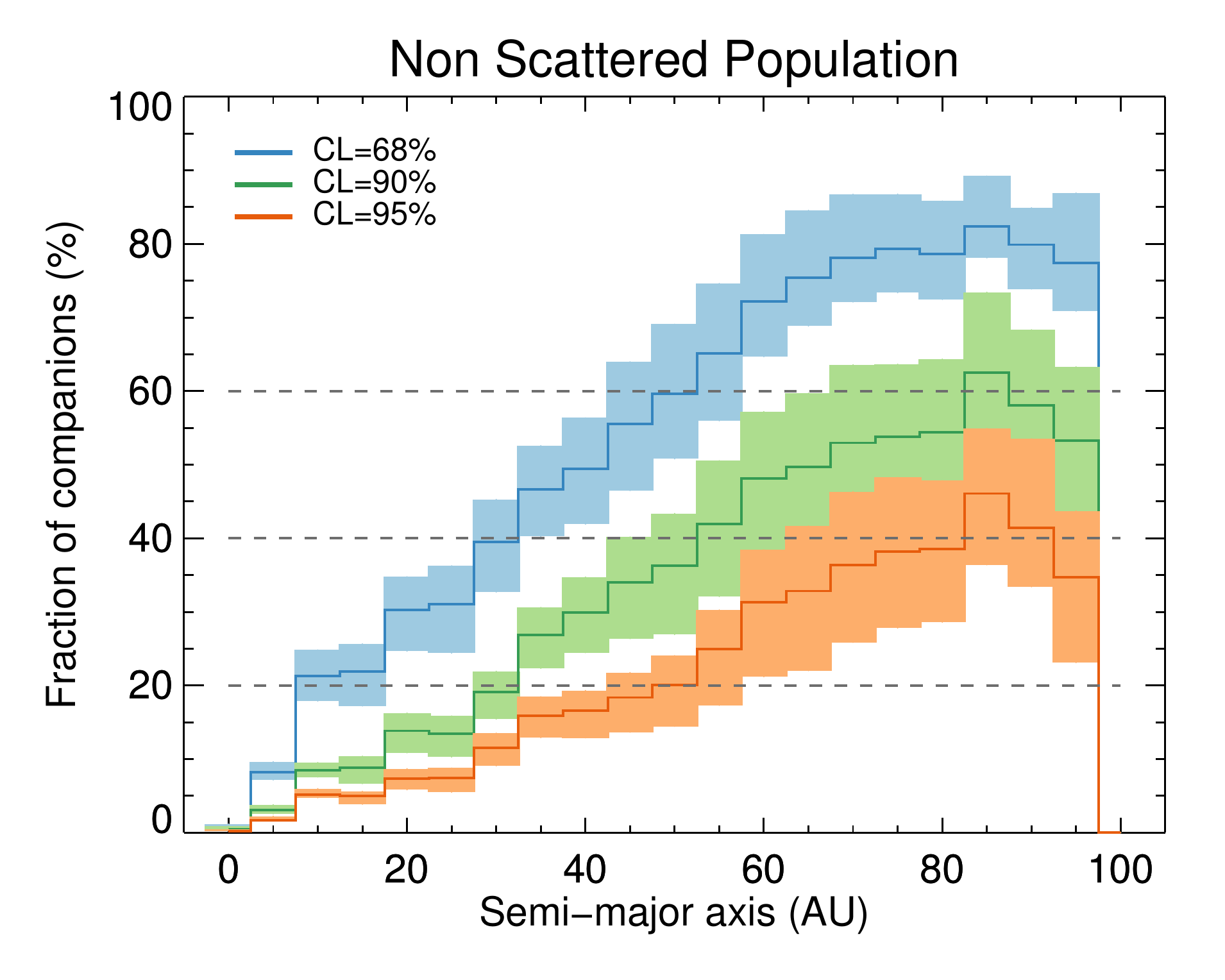}
    \includegraphics[width=0.48\textwidth]{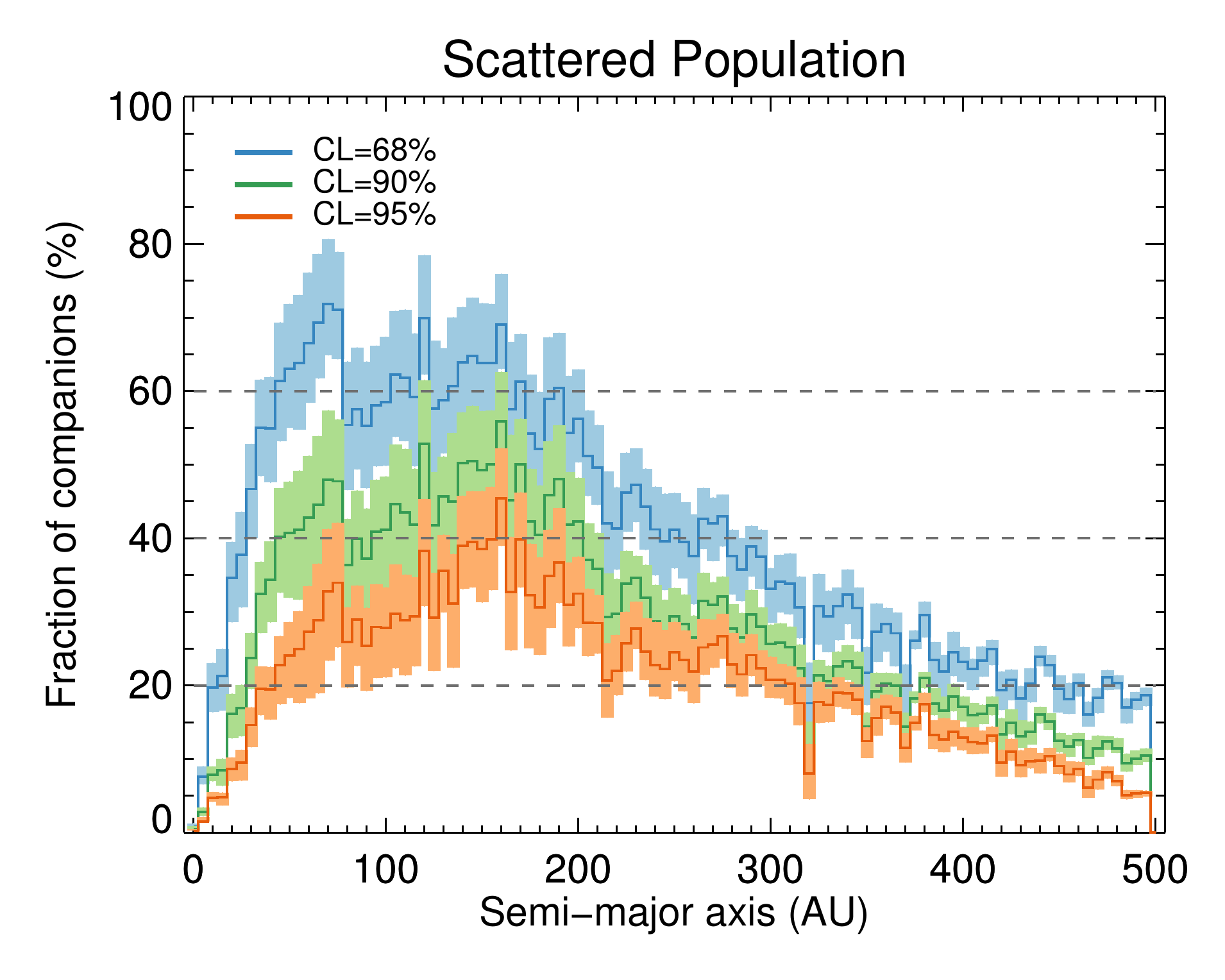}
    \caption{Histogram of the mass and separation of companions detectable at various confidence levels, for the scattered (right panels) and non-scattered (left panels) population. Similarly to Fig.~\ref{fig:cal_frac}, the plain lines represent the values for the best age estimate of the stars, and the colored envelopes show the variations observed when considering the minimum and maximum ages. The fraction of detected companions is smaller for the maximum ages and larger for the minimum ages. For the histograms in mass, semi-major axes from 1 to 500~AU are considered, and for the histograms in semi-major axis, masses from 1 to 75~\MJup are considered. A confidence level of 95\% means that for a given bin, the height of the bin represents the fraction of companions in that bin that have been detected in 95\% (or more) of the simulated surveys. The small drop at 330 AU in the scattered population is numerical.}
    \label{fig:mod_frac}
\end{figure*}

\begin{figure*}
    \centering
    \includegraphics[width=1.0\textwidth]{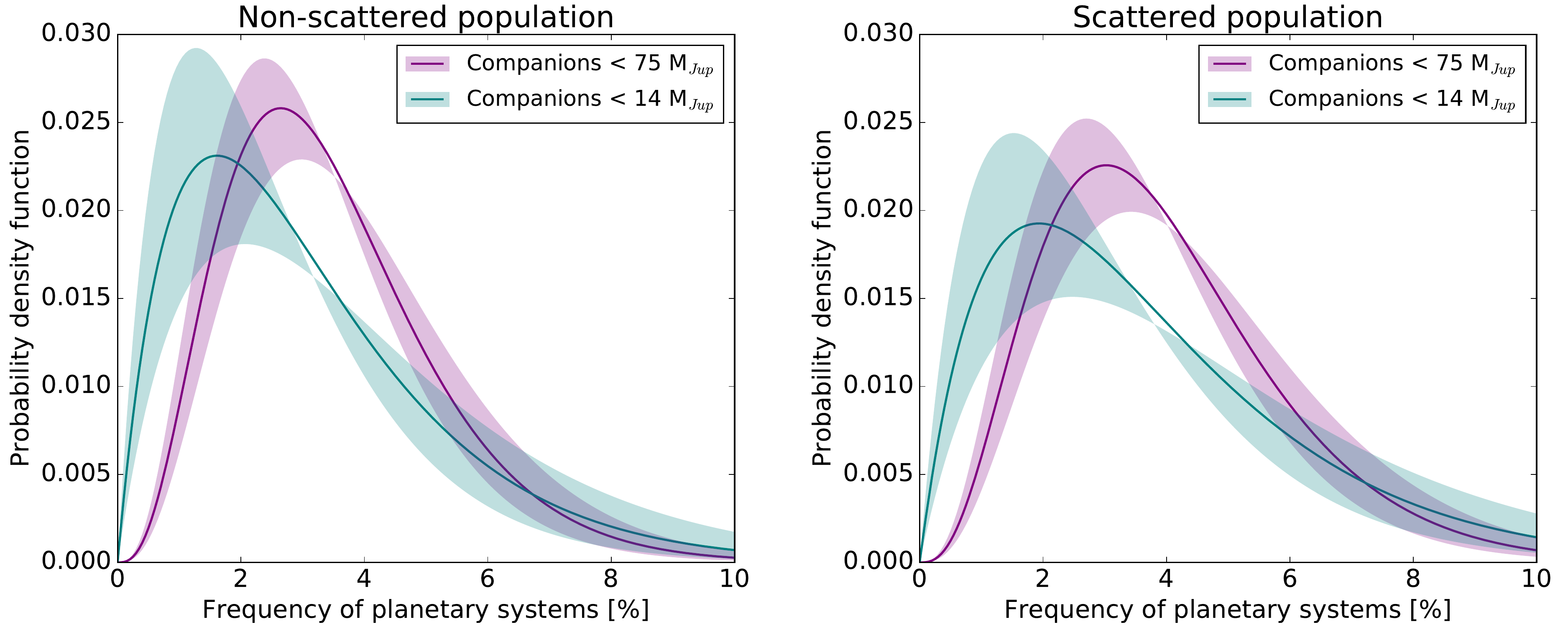}
    \caption{Probability distribution of the frequency of stars hosting at least one planetary system, based on the non-scattered (left) and scattered (right) populations described in Sec.~\ref{sec:planet_population_synthesis_models}. Similarly to Figs.~\ref{fig:cal_frac} and \ref{fig:mod_frac}, the plain lines represent the probability density function obtained when considering the best age estimate, and the shaded areas show the variations obtained when considering the full age range. The results are summarized in Table~\ref{tab:freq_pop}.}
    \label{fig:freq_GI}
\end{figure*}

One of the goals of our work is to test the predictions of the \citet{forgan2013} population synthesis model based on the GI formation mechanism and described in Sect.~\ref{sec:planet_population_synthesis_models}. The output of the GI models provides mass, physical separation, and eccentricity for all the planets in the population. We use this information to compare the non-scattered and scattered populations with the detection limits of the survey presented in Fig.~\ref{fig:1dlim}. Qualitatively, it appears that if a large population of planets formed by GI exists, it should be detectable from these observations.

To try to quantify this statement, we use the MESS code to assess the detectability of this GI population for our survey. The masses, semi-major axes, and eccentricities are taken directly from the generated population, and the code takes care of randomly generating all the missing orbital parameters needed to project the planetary orbit and evaluate the projected separation at the time of the observation. The detectability of the full population is then tested against the detection limits described in Sec.~\ref{sec:detection_limits}, individually for all the stars in the sample.

This process allows us to estimate the probability of detection of each companion in the population.\footnote{In this analysis, we ignore companions more massive than 75~\MJup in the population because close stellar binaries were removed from the statistical sample around each star in the sample.} Figure~\ref{fig:mod_frac} summarizes the result of the analysis: for each panel the fraction of companions with average detection probability greater than 68\%, 90\%, and 95\% are shown as a function of the companion mass (top) or separation (bottom) for both the scattered and non-scattered populations. The fraction of companions detected in each bin is of course highly correlated to the planets distribution in the original populations (Fig.~\ref{fig:sma_histo}). The apparent inconsistency between Figs.~\ref{fig:mod_cmp} and \ref{fig:mod_frac} is caused by the fact that Fig.~\ref{fig:mod_cmp} compares the populations to the mean probability, which is pessimistic in the case of stars with poor detection limits (see Fig.~\ref{fig:1dlim}), whereas Fig.~\ref{fig:mod_frac} is based on a per-star analysis which is then combined together to produce the figure.

The plots as a function of semi-major axis in Fig.~\ref{fig:mod_frac} show that more than 30\% of companions with masses in the 1--75~\MJup range and separations in the 70--200~AU range would be detected in 95\% of the simulated surveys, for both the scattered and non-scattered scenarios. This number even goes above 50\% when considering 68\% of the simulated surveys. In the plots as a function of mass, there is a significant loss of sensitivity towards small masses (<10~\MJup) because the population contains a large fraction of 1--20~\MJup planets to which we are not very sensitive. Nevertheless, the population also contains a significant fraction of 20--50~\MJup objects to which we are highly sensitive, both in the scattered and non-scattered scenarios. In other words, this analysis shows that under the hypothesis of the model, a population of GI-formed planets would have a great chance of being detected by our survey, even if scattering is negligible.

\begin{table}
    \caption[]{Fraction of stars that host a planetary system based on the GI populations described in Sect.~\ref{sec:planet_population_synthesis_models}.}
    \label{tab:freq_pop}
    \centering
    \begin{tabular}{ccccc}
    \hline \hline
    Mass               & $N_{det}$\tablefootmark{a} & $F_{best}$\tablefootmark{b} & \multicolumn{2}{c}{$[F_{min}, F_{max}]$\tablefootmark{c}} \\
    ($M_\mathrm{Jup}$) &                            & (\%)                        & CL=68\%      & CL=95\%       \\
    \hline
    \multicolumn{5}{c}{Non-scattered population} \\
    \hline
     $\le$75           & 3                          & 2.65                        & [1.85, 5.10] & [1.00, 7.50] \\
     $\le$14           & 1                          & 1.60                        & [1.15, 5.20] & [0.40, 8.70] \\
    \hline
    \multicolumn{5}{c}{Scattered population} \\
    \hline
     $\le$75           & 3                          & 3.00                        & [2.15, 5.85] & [1.10, 8.60] \\
     $\le$14           & 1                          & 1.95                        & [1.40, 6.20] & [0.50, 10.45] \\
    \hline
    \end{tabular} 
    \tablefoot{Numbers are provided for the best stellar ages. Values for the minimum and maximum ages are provided in Appendix~\ref{sec:app_freq}. \tablefoottext{a}{Number of detections in the considered mass range. \tablefoottext{b}{Best value of the planet frequency compatible with the observations.} \tablefoottext{c}{Minimum and maximum values of the frequency compatible with the results for a given confidence level (CL).}}}
\end{table}

The MESS code provided us with a probability of detecting each companion belonging to the GI populations around each star in our sample, taking into account the observed detection limits and the projection effects (see above). Such results can be used to constrain the value of the frequency companions compatible with our survey results, in a similar fashion to what is described in Sec.~\ref{sec:frequency_estimation}. Most companions in the populations are part of multiple systems (both in the non-scattered and scattered case, see Fig.~\ref{fig:sys_barplot}). We therefore decided that a more meaningful way to interpret our results would be to consider the probability of detecting a system with one or more companions, instead of considering each companion singularly. We therefore considered a synthetic system detected if at least one of its components had a detection probability greater than zero. In cases where more than one component satisfied such a requirement, then the highest value of the detection probability among the components was adopted as the detection probability of that system. Instead of limiting our analysis to a specific mass and semi-major axis range, we considered all the systems in the populations with companions up to 75~\MJup. Once the detection probability of each synthetic system was obtained for all the stars in the sample, the average over all the systems was adopted as the detection probability for each star. This allowed us to estimate the fraction of stars hosting at least one system consisting of one or more companions formed via GI, compatible with our observations. The results are shown in Fig.~\ref{fig:freq_GI} and summarized in Table~\ref{tab:freq_pop} for both the non-scattered and scattered case.

All but one companion (AB Pic b) detected in our survey have estimated masses considerably higher than 14~\MJup, therefore they are more properly characterized as brown dwarfs than planets. Thus, we have calculated limits on F$_{ps}$ both for planets and higher mass companions as well as for the planets only case (in this case, defining ``planet'' as anything with mass $\le$14~\MJup). When considering all companions (three detections), we estimate F$_{ps}^{ns}$ = 2.65$^{+4.85}_{-1.00}\%$ at the 95\% confidence level for the non-scattered population and F$_{ps}^{ws}$ = 3.00$^{+5.60}_{-1.90}\%$ for the scattered case. For companions with masses $\leq$~14\MJup (one detection), these estimations drop to F$_{ps}^{ns}$ = 1.60$^{+7.10}_{-1.20}\%$ and F$_{ps}^{ws}$ = 1.95$^{+8.50}_{-1.45}\%$ respectively.

It is interesting to note that the best estimates of F$_{ps}$ are very similar in the scattered and non-scattered case. This is likely the result of the shape of the semi-major axis distribution of the population: in Fig.~\ref{fig:sma_histo} we see that although the scattering sends many objects at very large orbital separations, the overall shape of the distribution remains dominated by a large peak in the 10--100~AU range where our observations are the most sensitive. The effect of the tail of the population that is scattered in the 100-400~AU range is almost negligible and has little influence on the F$_{ps}$ estimation. The main consequence is that if we assume that all the detected companions are the result of a GI-like formation scenario, then our observations cannot distinguish between the scattered and non-scattered scenarios: both scenarios appear equally likely when compared to the observations.

\section{Discussion and conclusions}
\label{sec:conclusions}

\begin{figure}
    \centering
    \includegraphics[width=0.5\textwidth]{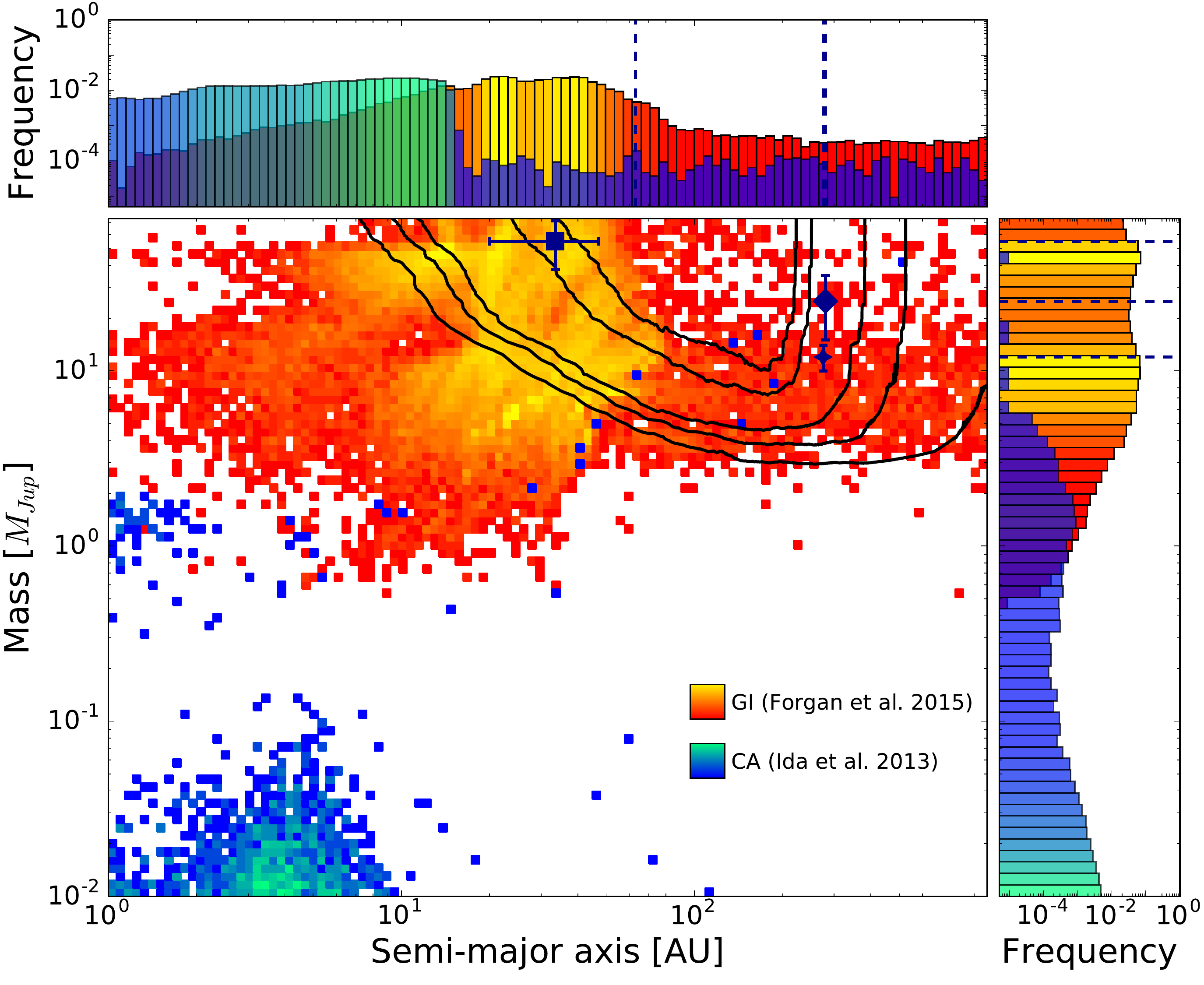}
    \caption{Density plots representing scattered populations based on GI \citep{forgan2013,forgan2015} and CA \citep{ida2013} compared to the detections in our sample and to the mean detection probabilities of the observations (contour lines are at 5\%, 25\%, 50\%, 90\%, and 95\%). Density colors go from dark blue (low occurrence) to cyan (high occurrence) for the CA population, and from red (low occurrence) to yellow (high occurrence) for the GI population. The histograms on top and on the right represent the relative frequency in each bin of semi-major axis and planetary mass respectively. The histograms take into account the whole population, including the planets that are outside of the visibility window of the main plot. The semi-major axis and mass of the known companions are represented with dashed lines in the histograms.}
    \label{fig:cmp_gi_ca}
\end{figure}

The dearth of detections of massive bodies at large radii challenges the gravitational instability theory of planet formation. This theory proposes that giant planets and brown dwarfs may be formed via disk fragmentation, which is predicted to occur at semi-major axes of at least 30--50~AU and larger. 

In a previous attempt at constraining the GI formation mechanism using the Gemini Deep Planet Survey sample \citep{lafreniere2007}, \citet{janson2012} used a toy model (Klahr et al., unpublished) to define the range of mass and semi-major axis where formation by GI would in principle be allowed around FGKM-type stars. In their approach, they consider that two criteria need to be satisfied simultaneously to form a planet by GI: the Toomre parameter \citep{toomre64} needs to be small enough for an instability to occur, which translates into a minimum mass of the resulting planet, and the cooling timescale of the disk needs to be short enough for fragmentation to occur \citep{gammie01}, which translates into a minimum semi-major axis of the resulting planet. The upper limit of F$_{ps}<10$\% with 98\% confidence derived by \citet{janson2012} is in very good agreement with our findings for the non-scattered population, based on a much more realistic model. We also note that \citet{rameau2013} reached similar conclusions as \citet{janson2012} regarding the rarity of planets formed by GI around A-stars using the same toy model approach.

Our study goes one step further and considers for the first time a more complete formation model based on GI that includes the scattering of the planets. The semi-major axis distribution of GI objects can be significantly modified by orbital migration \citep{Baruteau2011,Zhu2012}, dynamical interactions with neighboring GI objects, and nearby stellar systems \citep{forgan2015}. The masses of these bodies can also be reduced by Roche lobe overflow as they migrate inwards, drawing them into the parameter space more associated with core accretion \citep{nayakshin2010}. That being said, it is equally true that statistical studies of these processes produce populations with properties that do overlap with core accretion, but they necessarily produce a large fraction of objects which remain at large semi-major axis and with masses greater than a few Jupiter masses.

The fact that this high mass component, easily detectable, is not present in most of the observations presented here places strong constraints on how frequently disk fragmentation occurs. With only three detections in the brown-dwarf and planetary-mass regime, our analysis suggests that if GI is the predominant formation scenario for these objects then it is rare (<5\%). This result holds whether or not scattering is considered. We note that even though our observations are not sensitive to the very widest separations over which outward-scattered planets reside in the GI population, the Spitzer/IRAC direct imaging survey by \citet{durkan2016} has placed a frequency upper limit of 9\% for planets in the 0.5--13~\MJup and 100-1000~AU ranges, which is compatible with our results closer in.

Our conclusion that GI is rare is also consistent with an analysis by \citet{rice2015}, who considered the scattering and then tidal evolution of a population of planetary-mass bodies initially on wide orbits. Their analysis suggested that if GI were common, there should be a large population of ``hot'', or proto-``hot'', Jupiters that formed via GI on wide orbits, were scattered onto eccentric orbits by more distant stellar companions, and were then tidally circularized by interactions with their host star. That the population of known ``hot'', and proto-``hot'', Jupiters is inconsistent with this origin suggests that the formation of planetary-mass bodies by GI is rare.  

Even though GI cannot be common, it is predicting a mass distribution of wide-orbit sub-stellar companions that is much closer to what is observed than what CA scenarios predict. To illustrate this, we compare in Fig.~\ref{fig:cmp_gi_ca} two populations that include scattering: one based on the GI scenario, produced by the model of \citet{forgan2015}, and the other based on the CA scenario, produced by the model of \citet{ida2013}. While the bulk of the CA population is dominated by sub-Jupiter mass planets within $\sim$10 AU, there is an overlap with the GI population in the 1--10~\MJup range within 10~AU. However, the mass histogram clearly shows that GI dominates when considering masses above 5~\MJup and separations above 20~AU, similar to those of the imaged companions. CA cannot form massive brown dwarfs such as PZ~Tel~B, and even if extreme cases of scattered CA could form objects similar to TYC~8047-0232~B and AB~Pic~B, observing two cases of very large separation (>200 AU) massive companions while not observing the more common and easily detectable population that is predicted in the 50--200 AU range appears extremely improbable. Using a simple Monte Carlo simulation based on the mass and semi-major axis distributions of the two populations, and on the average detection probability map of the survey (Fig.~\ref{fig:mean_prob}), we estimate that it is $\sim$600\,000 times more likely to obtain three detections from the GI population than from the CA population. It seems therefore extremely unlikely for CA to be the preferential mechanism that explains the formation of the three imaged companions in the full NaCo-LP sample. This is of course even more the case when scattering is not included, since in that case the CA model produces no planets at separations wider than 20--30~AU.

We cannot completely rule out the formation of some of the observed companions by gravo-turbulent fragmentation \citep{hopkins2013}, that is, following a binary star-like formation scenario. \citet{reggiani2016} have showed that the NaCo-LP observations are compatible with a sub-stellar companion mass function constituted by a superposition of the stellar mass-ratio distribution and a planet mass function extrapolated from radial velocity detections. In that framework, our conclusions would still be valid because their model shows that to be compatible with the observations, they need to apply a cutoff of the RV extrapolation well below 100~AU, that is, typically in the range where CA still manages to produce a large fraction of planets. This value of the cutoff is actually supported by many previous direct imaging surveys \citep[e.g.,][]{lafreniere2007,kasper2007,chauvin2010}.

Our analysis of the planet frequency in Sect.~\ref{sec:frequency_estimation} assumes a linear flat distribution for the planet's mass and semi-major axis to derive the young giant planet frequency at large orbital separation. It can be argued that this approach, which does not rely on unknown or improperly modeled physics, is less prone to bias the estimation of the planet frequency. However, it puts an equal statistical weight on planets at all masses and separations, which is in contradiction with all models of planet formation, either based on CA \citep{mordasini2009,ida2004,ida2013} or GI \citep{forgan2013,forgan2015,nayakshin2016b}, which all agree that the formation of very massive companions at large orbital separation occurs much less frequently than less massive, closer-in planets. Although based on models, this information is important to take into account prior to the analysis. Despite the limitations highlighted at the end of Sect.~\ref{sec:planet_population_synthesis_models}, the models of \citet{forgan2013} constitute state-of-the-art models of formation by GI. It is important to be cautious in deriving quantitative estimations based on these models, but the fact that our frequency estimations in Sect.~\ref{sec:frequency_estimation} are compatible with the ones derived from the population synthesis models is a good sanity check.

It is also an essential first step for on-going large scale surveys with new-generation high-contrast imagers. These surveys, which are sensitive to lower masses at smaller orbital separations, are exploring the regime where CA and GI could have similar chances to form giant planets. In that context, statistical analysis work based on the outcomes of these surveys will need to make a stronger link between the observations and planet formation theories. This means focusing not only on deriving the occurrence rate of giant companions, but more importantly on deriving the properties of their underlying mass, separation, and eccentricity distributions. We have tried to initiate a first step in this direction using population synthesis models, and we hope that more work will soon follow.

\begin{acknowledgements}
The authors warmly thank B. Biller, T. Brandt, G. Chauvin, A. Heinze, M. Kasper, D. Lafreni\`ere, P. Lowrance, E. Masciadri, T. Meshkat, J. Rameau, I. Song and A. Vigan for providing the detection limits and candidates information for their surveys. We are also extremely grateful to S. Ida for providing planetary populations. Finally, AV is extremely grateful to \'Elodie Choquet, R\'emi Soummer, Laurent Pueyo and Brendan Hagan for fruitful discussions on the definition of a standard file format for high-contrast imaging surveys. \\

JLB, MB, GC, PD, JH, AML, CM, AV and AZ acknowledge support in France from the French National Research Agency (ANR) through project grants ANR10-BLANC0504-01 (GUEPARD) and ANR-14-CE33-0018 (GIPSE). DHF gratefully acknowledges support from the ECOGAL project, grant agreement 291227, funded by the European Research Council under ERC-2011-ADG. MB, SD, EC, RG, DM and SM acknowledge support from PRIN-INAF ``Planetary systems at young ages and the interactions with their active host stars'' and from the ``Progetti Premiali'' funding scheme of the Italian Ministry of Education, University, and Research. BB gratefully acknowledges support from STFC grant ST/M001229/1. JC was supported by the U.S. National Science Foundation under award no. 1009203. CM acknowledges the support from the Swiss National Science Foundation under grant BSSGI0$\_$155816 ``Planets In Time''. The research leading to these results has received funding from the European Union Seventh Framework Programme (FP7/2007-2013) under grant Agreement No. 313014 (ETAEARTH).
\end{acknowledgements}

\bibliographystyle{aa}
\bibliography{paper}

\appendix

\section{Full sample}
\label{sec:full_sample}

The names and properties of the targets considered in the full NaCo-LP sample are detailed in Table~\ref{tab:full_sample}. Different properties of the sample are plotted in Fig.~\ref{fig:sample_histograms}.

\onecolumn

\begin{landscape}
    \begin{longtable}{lr@{ }r@{ }lr@{ }r@{ }lccccccc>{$}c<{$}c}
    \caption{\label{tab:full_sample}Full NaCo-LP sample (observed + archive).}  \\
    \hline\hline
            Target & \multicolumn{3}{c}{$\alpha$}&\multicolumn{3}{c}{$\delta$}&SpT&Survey\tablefootmark{a} & MG\tablefootmark{b} &  Age & Age min &  Age max & Distance & \textrm{Fe/H} & Mass  \\
                   & \multicolumn{3}{c}{(J2000.0)}&\multicolumn{3}{c}{(J2000.0)}& &                        &                     &(Myr) &(Myr)&(Myr) & (pc) &      & (\MSun) \\
    \hline
    \endfirsthead
    \caption{continued.} \\
    \hline\hline
            Target & \multicolumn{3}{c}{$\alpha$}&\multicolumn{3}{c}{$\delta$}&SpT&Survey\tablefootmark{a} & MG\tablefootmark{b} & Best age & Age min & Age max & Distance & \textrm{Fe/H} & Mass  \\
                   & \multicolumn{3}{c}{(J2000.0)}&\multicolumn{3}{c}{(J2000.0)}& &                        &                     &(Myr) &(Myr)&(Myr) & (pc) &      & (\MSun) \\
    \hline
    \endhead
    \hline
    \endfoot

\object{HIP\,490}          & 00&05&52.50 & $-$41&45&11.00 & G0V    & \surv{L05}                                                        & Tuc/Hor    &     45    &     35    &     50    &    39.4    &   -0.06    &    1.1    \\
\object{HIP\,544}          & 00&06&36.78 & $+$29&01&17.41 & K0V    & \surv{L07},\surv{H10},\surv{R13},\surv{B14}                       & Her/Lyr    &    250    &    200    &    300    &    13.7    &    0.00    &    1.0    \\
\object{HIP\,560}          & 00&06&50.10 & $-$23&06&27.00 & F3V    & \surv{S06},\surv{B13}                                             & Beta Pic   &     24    &     19    &     29    &    39.4    &   -0.13    &    1.4    \\
\object{HIP\,682}          & 00&08&25.75 & $+$06&37&00.50 & G2V    & \surv{R13}                                                        &            &    200    &    120    &    350    &    39.1    &    0.11    &    1.1    \\
\object{HIP\,919}          & 00&11&22.44 & $+$30&26&58.47 & K0V    & \surv{L07}                                                        & Car?       &    250    &    200    &    300    &    34.2    &    0.20    &    1.0    \\
\object{HIP\,1113}         & 00&13&53.00 & $-$74&41&18.00 & G8V    & \surv{K07},\surv{C10},\surv{S06}                                  & Tuc/Hor    &     45    &     35    &     50    &    44.4    &   -0.06    &    0.9    \\
\object{HIP\,1134}         & 00&14&10.25 & $-$07&11&56.85 & F5     & \surv{S06},\surv{R13},\surv{B14}                                  & Col        &     42    &     35    &     50    &    47.1    &   -0.05    &    1.2    \\
\object{PW\,And}           & 00&18&20.89 & $+$30&57&22.23 & K2V    & \surv{L05},\surv{L07},\surv{H10}                                  & AB Dor     &    149    &    100    &    180    &    29.0    &   -0.01    &    0.8    \\
\object{HIP\,1481}         & 00&18&26.10 & $-$63&28&39.00 & F8V    & \surv{K07},\surv{B07},\surv{C10},\surv{S06},\surv{B13}            & Tuc/Hor    &     45    &     35    &     50    &    41.5    &   -0.06    &    1.2    \\
\object{HIP\,2729}         & 00&34&51.20 & $-$61&54&58.00 & K4Ve   & \surv{M05},\surv{K07},\surv{S06},\surv{B13}                       & Tuc/Hor    &     45    &     35    &     50    &    43.9    &   -0.06    &    0.9    \\
\object{QT\,And}           & 00&41&17.34 & $+$34&25&16.85 & K7Ve   & \surv{L05}                                                        &            &     80    &     50    &    150    &    46.9    &  \ldots    &    0.8    \\
\object{TYC\,9351-1110-1}  & 00&42&20.30 & $-$77&47&40.00 & K3Ve   & \surv{C10}                                                        & Tuc/Hor    &     45    &     35    &     50    &    49.4    &   -0.06    &    0.8    \\
\object{HIP\,3589}         & 00&45&50.89 & $+$54&58&40.18 & F8V    & \surv{S06},\surv{B14}                                             & AB Dor     &    149    &    100    &    180    &    52.5    &   -0.01    &    1.2    \\
\object{HIP\,4907}         & 01&02&57.22 & $+$69&13&37.42 & G5     & \surv{L07}                                                        &            &    800    &    500    &   1200    &    25.5    &   -0.14    &    0.9    \\
\object{HIP\,5191}         & 01&06&26.15 & $-$14&17&47.11 & K1V    & \surv{C10},\surv{S06},\surv{B13}                                  & AB Dor     &    149    &    100    &    180    &    47.3    &   -0.01    &    0.8    \\
\object{HIP\,6276}         & 01&20&32.27 & $-$11&28&03.74 & G9V    & \surv{S06},\surv{R13}                                             & AB Dor     &    149    &    100    &    180    &    34.4    &   -0.01    &    0.9    \\
\object{HIP\,6485}         & 01&23&21.30 & $-$57&28&51.00 & G7V    & \surv{B07},\surv{C10},\surv{S06}                                  & Tuc/Hor    &     45    &     35    &     50    &    49.5    &   -0.06    &    1.0    \\
\object{HIP\,6856}         & 01&28&08.70 & $-$52&38&19.00 & K1V    & \surv{B07},\surv{C10},\surv{S06}                                  & Tuc/Hor    &     45    &     35    &     50    &    36.0    &   -0.06    &    0.8    \\
\object{HIP\,7235}         & 01&33&15.81 & $-$24&10&40.66 & K0V    & \surv{L07}                                                        &            &   1300    &   1000    &   1600    &    19.1    &   -0.08    &    0.9    \\
\object{TYC\,1752-0063-1}  & 01&37&23.23 & $+$26&57&12.09 & K5Ve   & \surv{B14}                                                        & AB Dor?    &    149    &    100    &    180    &    60.7    &   -0.01    &    0.7    \\
\object{HIP\,7576}         & 01&37&35.47 & $-$06&45&37.52 & G5     & \surv{L07},\surv{R13}                                             & Her/Lyr    &    250    &    200    &    300    &    24.0    &    0.00    &    0.8    \\
\object{HIP\,7805}         & 01&40&24.07 & $-$60&59&56.63 & F2IV/V & \surv{C10},\surv{R13}                                             & LA?        &    200    &     30    &   1500    &    67.2    &   -0.18    &    1.3    \\
\object{HIP\,7978}         & 01&42&29.32 & $-$53&44&27.00 & F9     & \surv{R13}                                                        &            &   1500    &    700    &   2500    &    17.4    &    0.00    &    1.1    \\
\object{HIP\,8102}         & 01&44&04.08 & $-$15&56&14.93 & G8V    & \surv{H10}                                                        &            &   8000    &   7000    &  10000    &     3.7    &   -0.52    &    0.8    \\
\object{TYC\,8047-0232-1}  & 01&52&14.60 & $-$52&19&33.00 & K2V(e) & \surv{C10}                                                        & Tuc/Hor    &     42    &     35    &     50    &    80.3    &   -0.05    &    0.8    \\
\object{TYC\,8489-1155-1}  & 02&07&32.20 & $-$59&40&21.00 & K5Ve   & \surv{S06}                                                        & Tuc/Hor    &     45    &     35    &     50    &    44.2    &   -0.06    &    0.7    \\
\object{HIP\,10679}        & 02&17&24.75 & $+$28&44&30.44 & G2V    & \surv{S06},\surv{B14},\surv{B13}                                  & Beta Pic   &     24    &     19    &     29    &    32.8    &   -0.13    &    1.0    \\
\object{HIP\,10680}        & 02&17&25.29 & $+$28&44&42.16 & F5V    & \surv{S06}                                                        & Beta Pic   &     24    &     19    &     29    &    32.8    &   -0.13    &    1.1    \\
\object{HIP\,11360}        & 02&26&16.20 & $+$06&17&33.00 & F2     & \surv{C15},\surv{V12},\surv{R13},\surv{B13}                       & Tuc/Hor    &     45    &     35    &     50    &    45.2    &  \ldots    &    1.4    \\
\object{HIP\,11437}        & 02&27&29.25 & $+$30&58&24.61 & K8     & \surv{S06},\surv{B14}                                             & Beta Pic   &     24    &     19    &     29    &    40.0    &   -0.13    &    0.8    \\
\object{CD-53\,544}        & 02&41&46.83 & $-$52&59&52.36 & K6V    & \surv{B13}                                                        & Tuc/Hor    &     45    &     35    &     50    &    42.0    &  \ldots    &    0.6    \\
\object{HIP\,12926}        & 02&46&15.21 & $+$25&38&59.64 & K1IV   & \surv{L07}                                                        &            &   4300    &   2000    &   7000    &    25.9    &   -0.10    &    0.9    \\
\object{HIP\,13402}        & 02&52&32.13 & $-$12&46&10.97 & K1V    & \surv{L05},\surv{M05},\surv{B07},\surv{L07},\surv{H10}            &            &    150    &     70    &    300    &    10.4    &    0.06    &    0.9    \\
\object{TYC\,8862-0019-1}  & 02&58&04.00 & $-$62&41&14.00 & K3Ve   & \surv{C10}                                                        & Car?       &     45    &     10    &     55    &    93.3    &   -0.07    &    0.8    \\
\object{HIP\,14150}        & 03&02&26.03 & $+$26&36&33.26 & G8V    & \surv{L07}                                                        &            &   4400    &   2000    &   8000    &    20.6    &    0.09    &    1.0    \\
\object{HIP\,14684}        & 03&09&42.29 & $-$09&34&46.59 & G0     & \surv{B13},\surv{C15}                                             & AB Dor     &    149    &    100    &    180    &    37.4    &   -0.01    &    0.9    \\
\object{HIP\,15323}        & 03&17&40.05 & $+$31&07&37.37 & G0     & \surv{L07}                                                        &            &    400    &    250    &    650    &    26.7    &    0.08    &    1.2    \\
\object{2E\,759}           & 03&20&49.50 & $-$19&16&10.00 & K7V    & \surv{L07}                                                        &            &    100    &     40    &    150    &    41.7    &  \ldots    &    0.7    \\
\object{TYC\,8060-1673-1}  & 03&30&49.10 & $-$45&55&57.00 & K3V    & \surv{C15}                                                        & Tuc/Hor    &     45    &     35    &     50    &    40.4    &   -0.06    &    0.8    \\
\object{HIP\,16537}        & 03&32&55.84 & $-$09&27&29.74 & K2V    & \surv{B07},\surv{L07},\surv{H10}                                  &            &    600    &    400    &    800    &     3.2    &   -0.11    &    0.8    \\
\object{HIP\,16563}        & 03&33&13.49 & $+$46&15&26.54 & G5     & \surv{B07},\surv{S06}                                             & AB Dor     &    149    &    100    &    180    &    34.4    &   -0.01    &    1.0    \\
\object{HD\,25284}         & 04&00&03.78 & $-$29&02&16.40 & K6V    & \surv{C10},\surv{S06}                                             & Tuc/Hor    &     45    &     35    &     50    &    49.3    &  \ldots    &    0.8    \\
\object{TYC\,5882-1169-1}  & 04&02&16.50 & $-$15&21&30.00 & K3/4   & \surv{S06}                                                        & Col        &     42    &     35    &     50    &    53.2    &   -0.05    &    0.8    \\
\object{HIP\,18859}        & 04&02&36.74 & $+$00&16&08.12 & F5V    & \surv{S06},\surv{L07},\surv{H10},\surv{B13}                       & AB Dor     &    149    &    100    &    180    &    18.8    &   -0.01    &    1.2    \\
\object{HIP\,19335}        & 04&08&36.62 & $+$38&02&23.05 & F7V    & \surv{S06}                                                        &            &    300    &    200    &    500    &    21.0    &    0.06    &    1.3    \\
\object{HIP\,19775}        & 04&14&22.60 & $-$38&19&02.00 & G3V    & \surv{C15}                                                        & Col        &     42    &     35    &     50    &    80.5    &   -0.05    &    1.1    \\
\object{HIP\,21632}        & 04&38&43.90 & $-$27&02&02.00 & G3V    & \surv{S06}                                                        & Tuc/Hor    &     45    &     35    &     50    &    56.2    &   -0.06    &    1.0    \\
\object{HIP\,21818}        & 04&41&18.86 & $+$20&54&05.44 & K3V    & \surv{L05},\surv{B07}                                             & UMa?       &    120    &     50    &    200    &    13.2    &    0.01    &    0.7    \\
\object{HIP\,22152}        & 04&46&00.58 & $+$76&36&39.76 & F7V    & \surv{S06}                                                        &            &    250    &    125    &    400    &    32.3    &   -0.14    &    1.2    \\
\object{HIP\,22226}        & 04&46&49.50 & $-$26&18&00.00 & F3V    & \surv{V12},\surv{R13}                                             & Col        &     42    &     35    &     50    &    80.3    &   -0.05    &    1.4    \\
\object{HIP\,22295}        & 04&48&05.18 & $-$80&46&45.30 & F7V    & \surv{R13}                                                        & Tuc/Hor    &     45    &     35    &     50    &    61.0    &   -0.06    &    1.2    \\
\object{HIP\,22449}        & 04&49&50.41 & $+$06&57&40.59 & F6V    & \surv{L07}                                                        &            &   1200    &    500    &   2000    &     8.1    &    0.00    &    1.3    \\
\object{HIP\,22844}        & 04&54&53.04 & $-$58&32&51.55 & F3V    & \surv{S06}                                                        &            &    250    &    100    &    800    &    30.4    &   -0.13    &    1.3    \\
\object{HIP\,23316}        & 05&00&51.90 & $-$41&01&07.00 & G5V    & \surv{C15}                                                        & Col        &     42    &     35    &     50    &    76.3    &   -0.05    &    1.0    \\
\object{HD\,32981}         & 05&06&27.70 & $-$15&49&30.00 & F8V    & \surv{C15}                                                        & AB Dor     &    149    &    100    &    180    &    86.7    &   -0.01    &    1.1    \\
\object{BD\,-09\,1108}     & 05&15&36.50 & $-$09&30&51.00 & G5     & \surv{C15}                                                        & Tuc/Hor    &     45    &     35    &     50    &    93.6    &   -0.06    &    1.0    \\
\object{HIP\,24947}        & 05&20&38.00 & $-$39&45&18.00 & F6V    & \surv{S06},\surv{B13}                                             & Col        &     42    &     35    &     50    &    48.3    &   -0.05    &    1.3    \\
\object{HIP\,25283}        & 05&24&30.17 & $-$38&58&10.76 & K6V    & \surv{S06},\surv{C10},\surv{B13}                                  & AB Dor     &    149    &    100    &    180    &    18.0    &   -0.01    &    0.7    \\
\object{HIP\,25434}        & 05&26&22.98 & $-$43&22&36.38 & G0     & \surv{C15}                                                        & Col        &     42    &     35    &     50    &    79.1    &  \ldots    &    1.2    \\
\object{TYC\,9162-0698-1}  & 05&29&27.10 & $-$68&52&05.00 & G6V    & \surv{C15}                                                        & Col?       &     42    &     35    &     50    &    98.7    &  \ldots    &    1.1    \\
\object{HD\,36869}         & 05&34&09.16 & $-$15&17&03.18 & G2V    & \surv{B14}                                                        & Col        &     42    &     35    &     50    &    57.7    &   -0.05    &    1.0    \\
\object{HIP\,26369}        & 05&36&55.07 & $-$47&57&47.99 & K6Ve   & \surv{S06},\surv{C10},\surv{B13}                                  & AB Dor     &    149    &    100    &    180    &    25.1    &   -0.01    &    0.7    \\
\object{HIP\,26373}        & 05&36&56.85 & $-$47&57&52.87 & K0V    & \surv{B07},\surv{C10},\surv{S06},\surv{B13}                       & AB Dor     &    149    &    100    &    180    &    25.1    &   -0.01    &    0.9    \\
\object{HIP\,26453}        & 05&37&39.63 & $-$28&37&34.66 & F3V    & \surv{L05},\surv{R13}                                             & Col        &     42    &     35    &     50    &    56.8    &   -0.05    &    1.4    \\
\object{TYC\,5346-132-1}   & 05&38&35.00 & $-$08&56&04.00 & G7     & \surv{C15}                                                        & Col        &     42    &     35    &     50    &    81.2    &   -0.05    &    1.0    \\
\object{HIP\,26779}        & 05&41&20.34 & $+$53&28&51.81 & K1V    & \surv{H10}                                                        &            &    600    &    400    &    800    &    12.3    &    0.16    &    0.9    \\
\object{HIP\,26990}        & 05&43&35.81 & $-$39&55&24.72 & G0V    & \surv{S06}                                                        & Col        &     42    &     35    &     50    &    55.4    &   -0.05    &    1.1    \\
\object{HIP\,27072}        & 05&44&27.79 & $-$22&26&54.18 & F7V    & \surv{H10}                                                        & UMa        &    450    &    380    &    550    &     8.9    &   -0.03    &    1.3    \\
\object{HIP\,29067}        & 06&07&55.25 & $+$67&58&36.55 & K8V    & \surv{B14}                                                        & Cas        &    400    &    200    &   1000    &    24.5    &    0.00    &    0.7    \\
\object{HIP\,29964}        & 06&18&28.20 & $-$72&02&41.00 & K4Ve   & \surv{M05},\surv{B07},\surv{K07},\surv{B13}                       & Beta Pic   &     24    &     19    &     29    &    38.6    &   -0.13    &    0.8    \\
\object{HIP\,30030}        & 06&19&08.10 & $-$03&26&20.00 & G0     & \surv{B07},\surv{S06},\surv{R13},\surv{B14}                       & Tuc/Hor    &     45    &     35    &     50    &    49.2    &   -0.06    &    1.1    \\
\object{HIP\,30034}        & 06&19&12.90 & $-$58&03&16.00 & K1V(e) & \surv{B07},\surv{C10},\surv{R13},\surv{B13}                       & Car        &     45    &     35    &     55    &    46.1    &   -0.07    &    0.9    \\
\object{HIP\,30261}        & 06&21&57.30 & $-$34&30&44.00 & G6V    & \surv{C15}                                                        &            &    120    &     70    &    200    &    61.8    &  \ldots    &    1.0    \\
\object{HIP\,30314}        & 06&22&30.94 & $-$60&13&07.15 & G1V    & \surv{B07},\surv{C10},\surv{R13},\surv{S06}                       & AB Dor     &    149    &    100    &    180    &    23.8    &   -0.01    &    1.1    \\
\object{TYC\,7617-0549-1}  & 06&26&06.90 & $-$41&02&54.00 & K0V    & \surv{C15}                                                        & Col        &     42    &     35    &     50    &    77.8    &   -0.05    &    0.9    \\
\object{HIP\,32235}        & 06&43&46.20 & $-$71&58&35.00 & G6V    & \surv{C15}                                                        & CAR        &     45    &     35    &     55    &    58.2    &   -0.07    &    1.0    \\
\object{TYC\,1355-214-1}   & 07&23&43.60 & $+$20&24&59.00 & K3     & \surv{L05},\surv{H10},\surv{B14}                                  & AB Dor     &    149    &    100    &    180    &    28.3    &   -0.01    &    0.7    \\
\object{TYC\,8128-1946-1}  & 07&28&22.00 & $-$49&08&38.00 & G8V    & \surv{C15}                                                        & Arg        &     50    &     40    &     70    &    89.7    &   -0.03    &    0.9    \\
\object{HIP\,36948}        & 07&35&47.46 & $-$32&12&14.04 & G3/G5V & \surv{C15}                                                        & Arg        &     50    &     40    &     70    &    35.3    &   -0.03    &    1.0    \\
\object{HIP\,37349}        & 07&39&59.33 & $-$03&35&51.03 & K2V    & \surv{H10}                                                        &            &    500    &    250    &    700    &    14.2    &    0.07    &    0.8    \\
\object{HIP\,37563}        & 07&42&36.06 & $-$59&17&50.80 & G3V    & \surv{C15}                                                        & Car-Near   &    250    &    200    &    300    &    32.8    &    0.08    &    1.1    \\
\object{HIP\,37923}        & 07&46&16.97 & $-$59&48&34.20 & K0V    & \surv{C15}                                                        & Car-Near   &    250    &    200    &    300    &    36.8    &    0.08    &    1.0    \\
\object{HIP\,40774}        & 08&19&19.05 & $+$01&20&19.91 & G5     & \surv{B14}                                                        & Car-Near?  &    300    &    200    &    500    &    22.9    &   -0.22    &    0.8    \\
\object{BD\,+02\,1951}     & 08&22&49.95 & $+$01&51&33.55 & G6V    & \surv{L05}                                                        & Her/Lyr    &    250    &    200    &    300    &    62.4    &    0.00    &    1.1    \\
\object{HIP\,41889}        & 08&32&30.53 & $+$15&49&26.19 & K5V    & \surv{S06}                                                        &            &    150    &     50    &    200    &    34.2    &  \ldots    &    0.7    \\
\object{HIP\,42438}        & 08&39&11.70 & $+$65&01&15.26 & G1.5Vb & \surv{L05},\surv{B07},\surv{L07},\surv{H10}                       & UMa        &    450    &    380    &    550    &    14.4    &   -0.03    &    1.0    \\
\object{HIP\,43410}        & 08&50&32.22 & $+$33&17&06.19 & F7Vn   & \surv{L07}                                                        &            &    300    &    200    &    450    &    28.4    &    0.05    &    1.3    \\
\object{HIP\,44526}        & 09&04&20.69 & $-$15&54&51.40 & K2V    & \surv{B14}                                                        & Cas        &    350    &    200    &    500    &    28.3    &    0.00    &    0.8    \\
\object{HD\,78141}         & 09&07&18.08 & $+$22&52&21.57 & K0     & \surv{L07},\surv{H10}                                             &            &    180    &    120    &    300    &    24.1    &  \ldots    &    0.8    \\
\object{HIP\,46580}        & 09&29&54.82 & $+$05&39&18.48 & K3V    & \surv{H10}                                                        &            &    600    &    400    &    800    &    12.9    &   -0.04    &    0.8    \\
\object{HIP\,46634}        & 09&30&35.00 & $+$10&35&59.83 & G5     & \surv{C15}                                                        &            &    350    &    150    &    600    &    42.3    &   -0.01    &    0.9    \\
\object{HIP\,46843}        & 09&32&43.76 & $+$26&59&18.71 & K0     & \surv{L05},\surv{B07},\surv{L07},\surv{H10},\surv{B13},\surv{B14} & Her/Lyr?   &    250    &    100    &    300    &    17.8    &   -0.03    &    0.9    \\
\object{HIP\,47646}        & 09&42&50.81 & $-$22&57&55.77 & F5V    & \surv{C15}                                                        &            &   1150    &    100    &   4000    &    73.6    &    0.04    &    1.3    \\
\object{TWA\,21}           & 10&13&14.80 & $-$52&30&54.00 & K3Ve   & \surv{C15}                                                        &            &     17    &      8    &     50    &    54.8    &  \ldots    &    1.0    \\
\object{HIP\,50660}        & 10&20&45.92 & $+$32&23&54.31 & K0V    & \surv{B14}                                                        &            &   1300    &    700    &   3000    &    47.1    &  \ldots    &    0.9    \\
\object{TYC\,6069-1214-1}  & 10&27&37.30 & $-$20&27&11.00 & K0V    & \surv{C15}                                                        &            &    100    &     45    &    150    &    67.8    &  \ldots    &    0.9    \\
\object{HIP\,51386}        & 10&29&42.23 & $+$01&29&28.02 & F5     & \surv{L07},\surv{C10},\surv{S06}                                  &            &    170    &    100    &    350    &    31.5    &    0.03    &    1.2    \\
\object{TYC\,7722-0207-1}  & 10&32&43.90 & $-$44&40&56.00 & K0V    & \surv{C15}                                                        &            &    130    &     70    &    200    &    65.8    &  \ldots    &    0.9    \\
\object{HIP\,51931}        & 10&36&30.79 & $-$13&50&35.82 & K2V    & \surv{L07}                                                        &            &    750    &    500    &   1000    &    32.2    &   -0.09    &    0.8    \\
\object{HIP\,52462}        & 10&43&28.27 & $-$29&03&51.43 & K1V    & \surv{B07},\surv{L07},\surv{B13}                                  &            &    170    &    100    &    300    &    21.4    &    0.03    &    0.9    \\
\object{HIP\,52787}        & 10&47&31.16 & $-$22&20&52.93 & K0V    & \surv{L07}                                                        &            &    250    &    100    &    400    &    34.5    &   -0.11    &    0.9    \\
\object{HIP\,53486}        & 10&56&30.80 & $+$07&23&18.60 & K0     & \surv{B14}                                                        & Cas        &    500    &    300    &    700    &    17.3    &    0.00    &    0.9    \\
\object{HIP\,54155}        & 11&04&41.47 & $-$04&13&15.92 & G5     & \surv{L07},\surv{H10},\surv{B14}                                  & Her/Lyr?   &    250    &    200    &    400    &    26.3    &   -0.01    &    0.9    \\
\object{HIP\,54745}        & 11&12&32.35 & $+$35&48&50.69 & G0V    & \surv{L07},\surv{B07}                                             &            &    500    &    300    &    750    &    21.9    &    0.08    &    1.1    \\
\object{HIP\,56445}        & 11&34&21.95 & $+$03&03&36.60 & F5V    & \surv{S06}                                                        &            &    900    &    600    &   1500    &    27.2    &   -0.01    &    1.3    \\
\object{HIP\,57370}        & 11&45&42.29 & $+$02&49&17.34 & K0     & \surv{L07}                                                        &            &   1000    &    600    &   1500    &    29.6    &    0.05    &    0.9    \\
\object{TWA\,19A}          & 11&47&24.54 & $-$49&53&03.00 & G4V    & \surv{C10}                                                        & LCC        &     16    &     10    &     22    &    91.7    &   -0.06    &    1.2    \\
\object{HIP\,58240}        & 11&56&42.31 & $-$32&16&05.40 & G3V    & \surv{C15}                                                        & Car-Near   &    250    &    200    &    300    &    31.8    &    0.08    &    1.0    \\
\object{HIP\,59280}        & 12&09&37.26 & $+$40&15&07.40 & K0V    & \surv{L07},\surv{B14}                                             &            &   1300    &   1000    &   1600    &    24.5    &    0.14    &    1.0    \\
\object{HIP\,60074}        & 12&19&06.50 & $+$16&32&53.87 & G2V    & \surv{L07}                                                        &            &    200    &    120    &    300    &    27.5    &   -0.05    &    1.1    \\
\object{TYC\,8238-1462-1}  & 12&21&55.70 & $-$49&46&12.00 & G9V    & \surv{C10}                                                        & LCC?       &     16    &     10    &     50    &    73.5    &  \ldots    &    0.9    \\
\object{TYC\,8979-1683-1}  & 12&28&25.40 & $-$63&20&59.00 & G7V    & \surv{C15}                                                        & LCC?       &     16    &      6    &     30    &    75.6    &  \ldots    &    1.2    \\
\object{HD\,108767B}       & 12&29&50.91 & $-$16&31&14.99 & K0V    & \surv{L07}                                                        &            &    180    &    100    &    350    &    26.6    &   -0.01    &    0.8    \\
\object{HIP\,61174}        & 12&32&04.23 & $-$16&11&45.63 & F2V    & \surv{L07}                                                        &            &   1300    &    900    &   1800    &    18.3    &   -0.02    &    1.4    \\
\object{TYC\,8992-0605-1}  & 12&36&39.00 & $-$63&44&43.00 & K2V    & \surv{C10}                                                        & LCC?       &     10    &      5    &     20    &    60.6    &  \ldots    &    1.0    \\
\object{BD\,+60\,1417}     & 12&43&33.27 & $+$60&00&52.66 & K0     & \surv{L07},\surv{H10}                                             &            &    200    &    100    &    350    &    48.8    &  \ldots    &    0.9    \\
\object{HIP\,62523}        & 12&48&47.05 & $+$24&50&24.81 & G7V    & \surv{L07}                                                        &            &   1100    &    700    &   1700    &    16.9    &    0.06    &    1.0    \\
\object{TYC\,9245-0617-1}  & 12&58&25.60 & $-$70&28&49.00 & K0Ve   & \surv{C15}                                                        & Cha        &     10    &      4    &     20    &    93.0    &    0.01    &    1.2    \\
\object{HIP\,63317}        & 12&58&31.97 & $+$38&16&43.55 & G5V    & \surv{B14}                                                        &            &    250    &    150    &    450    &    44.2    &  \ldots    &    1.0    \\
\object{HIP\,63862}        & 13&05&16.81 & $-$50&51&23.82 & G5V    & \surv{C15}                                                        &            &    180    &    120    &    250    &    49.0    &  \ldots    &    1.1    \\
\object{HIP\,64408}        & 13&12&03.20 & $-$37&48&11.00 & G3V    & \surv{B07}                                                        &            &   4400    &   3300    &   5400    &    20.7    &    0.19    &    1.3    \\
\object{HIP\,64797}        & 13&16&51.05 & $+$17&01&01.86 & K2V    & \surv{H10}                                                        &            &   1000    &    500    &   1500    &    11.1    &   -0.22    &    0.8    \\
\object{HIP\,65515}        & 13&25&45.53 & $+$56&58&13.78 & G9IV-V & \surv{L07}                                                        & LA?        &    350    &    250    &    500    &    21.6    &    0.08    &    0.9    \\
\object{TYC\,3460-0416-1}  & 13&27&12.11 & $+$45&58&26.36 & K7     & \surv{B14}                                                        & AB Dor?    &    149    &    100    &    200    &    70.2    &   -0.01    &    0.7    \\
\object{TYC\,7796-2110-1}  & 13&34&31.90 & $-$42&09&31.00 & K2IVe  & \surv{C15}                                                        & LCC?       &     17    &      6    &     60    &    92.1    &  \ldots    &    0.9    \\
\object{HIP\,66252}        & 13&34&43.21 & $-$08&20&31.33 & K7V    & \surv{L05},\surv{L07},\surv{B14}                                  &            &    150    &    100    &    200    &    20.2    &   -0.17    &    0.7    \\
\object{HIP\,67412}        & 13&48&58.19 & $-$01&35&34.64 & K0     & \surv{B14}                                                        &            &   1300    &    800    &   2500    &    37.7    &    0.03    &    0.9    \\
\object{HIP\,69357}        & 14&11&46.17 & $-$12&36&42.36 & K1V    & \surv{L07}                                                        &            &   1500    &    500    &   2000    &    21.6    &   -0.17    &    0.8    \\
\object{HIP\,69713B}       & 14&16&12.18 & $+$51&22&34.56 & K1V    & \surv{L07}                                                        &            &    400    &    250    &    550    &    29.1    &  \ldots    &    0.9    \\
\object{HIP\,71395}        & 14&36&00.56 & $+$09&44&47.47 & K0     & \surv{K07},\surv{B07},\surv{H10}                                  & UMa?       &    450    &    350    &    550    &    16.5    &    0.01    &    0.8    \\
\object{SAO\,252852}       & 14&42&28.61 & $-$64&58&41.40 & K5V    & \surv{M05}                                                        &            &    980    &    800    &   1160    &    16.6    &   -0.12    &    0.7    \\
\object{HIP\,71933}        & 14&42&43.60 & $-$48&47&59.00 & F8V    & \surv{C15}                                                        & UCL?       &     17    &      6    &    100    &    83.9    &  \ldots    &    1.3    \\
\object{HIP\,72146}        & 14&45&24.18 & $+$13&50&46.73 & K0     & \surv{L07}                                                        &            &   3500    &   2000    &   6000    &    18.9    &   -0.24    &    0.7    \\
\object{HIP\,72339}        & 14&47&32.73 & $+$00&16&53.31 & K0III  & \surv{L07}                                                        &            &   3500    &   1500    &   6000    &    31.8    &   -0.02    &    0.9    \\
\object{HIP\,72567}        & 14&50&15.81 & $+$23&54&42.64 & G2V    & \surv{L07},\surv{H10}                                             &            &    550    &    400    &    800    &    18.2    &    0.00    &    1.1    \\
\object{HIP\,73996}        & 15&07&18.07 & $+$24&52&09.10 & F5V    & \surv{B14}                                                        &            &   1000    &    400    &   1700    &    19.6    &    0.05    &    1.3    \\
\object{HIP\,74405}        & 15&12&23.40 & $-$75&15&16.00 & G9V    & \surv{C10}                                                        & Arg        &     50    &     40    &     70    &    50.3    &   -0.03    &    0.9    \\
\object{HIP\,75829}        & 15&29&23.59 & $+$80&27&00.96 & G5     & \surv{L07},\surv{H10}                                             & Her/Lyr?   &    250    &    150    &    350    &    21.5    &    0.09    &    0.9    \\
\object{HIP\,76829}        & 15&41&11.38 & $-$44&39&40.34 & F5IV-V & \surv{B13},\surv{C15}                                             & Her/Lyr    &    250    &    200    &   1000    &    17.4    &    0.00    &    1.4    \\
\object{TYC\,6781-0415-1}  & 15&41&31.20 & $-$25&20&36.00 & G9IVe  & \surv{C15}                                                        & US?        &     11    &      4    &     15    &   106.0    &  \ldots    &    1.6    \\
\object{HIP\,77199}        & 15&45&47.60 & $-$30&20&55.74 & K2V    & \surv{B07},\surv{S06}                                             &            &     12    &      6    &     20    &    40.4    &  \ldots    &    0.9    \\
\object{HIP\,77408}        & 15&48&09.46 & $+$01&34&18.26 & G8V    & \surv{L07}                                                        &            &    550    &    300    &   1000    &    21.3    &   -0.06    &    0.9    \\
\object{TYC\,7846-1538-1}  & 15&53&27.30 & $-$42&16&01.00 & G2     & \surv{C10},\surv{S06}                                             & UCL?       &     17    &     10    &     50    &    43.8    &  \ldots    &    1.1    \\
\object{TYC\,6209-0769-1}  & 16&14&13.80 & $-$19&39&36.00 & K0IV   & \surv{C15}                                                        &            &    120    &     50    &    200    &    43.9    &  \ldots    &    0.8    \\
\object{HIP\,80758}        & 16&29&20.20 & $-$30&57&40.00 & G9V(e) & \surv{C15}                                                        &            &     20    &     10    &     50    &    98.2    &  \ldots    &    1.1    \\
\object{HIP\,83988}        & 17&10&10.51 & $+$54&29&39.81 & K0     & \surv{H10}                                                        & UMa?       &    450    &    380    &    700    &    19.9    &   -0.20    &    0.7    \\
\object{HIP\,83996}        & 17&10&12.36 & $+$54&29&24.50 & K8     & \surv{H10}                                                        & UMa?       &    450    &    380    &    700    &    19.9    &   -0.20    &    0.7    \\
\object{TYC\,7362-0724-1}  & 17&16&58.00 & $-$31&09&04.00 & G5V    & \surv{C15}                                                        &            &     20    &     10    &     60    &    90.0    &  \ldots    &    1.0    \\
\object{BD\,-13\,4687}     & 17&37&46.47 & $-$13&14&46.64 & K3     & \surv{S06}                                                        & AB Dor     &    149    &    100    &    180    &    26.2    &   -0.01    &    0.1    \\
\object{HIP\,86672}        & 17&42&30.40 & $-$28&44&56.00 & G5V    & \surv{C15}                                                        &            &     40    &     20    &     80    &    78.0    &  \ldots    &    1.1    \\
\object{HIP\,87579}        & 17&53&29.94 & $+$21&19&31.04 & K0     & \surv{B14}                                                        & Cas?       &    500    &    350    &    700    &    24.4    &   -0.22    &    0.8    \\
\object{HIP\,87768}        & 17&55&44.90 & $+$18&30&01.41 & K5V    & \surv{B14}                                                        &            &    500    &    250    &    800    &    25.0    &    0.23    &    0.8    \\
\object{HIP\,88399}        & 18&03&03.40 & $-$51&38&56.00 & F6V    & \surv{C10},\surv{S06},\surv{R13}                                  & Beta Pic   &     24    &     19    &     29    &    48.1    &   -0.13    &    1.3    \\
\object{HIP\,88945}        & 18&09&21.38 & $+$29&57&06.16 & G0     & \surv{B07}                                                        & Cas?       &   1000    &    100    &   2000    &    24.8    &    0.01    &    1.0    \\
\object{HIP\,89829}        & 18&19&52.20 & $-$29&16&33.00 & G5V    & \surv{C15}                                                        & Beta Pic   &     24    &     19    &     29    &    72.6    &   -0.13    &    1.2    \\
\object{HIP\,91043}        & 18&34&20.10 & $+$18&41&24.23 & G0V    & \surv{B14}                                                        &            &     45    &     20    &     80    &    38.0    &    0.01    &    1.1    \\
\object{HIP\,92680}        & 18&53&05.90 & $-$50&10&50.00 & G9IV   & \surv{M05},\surv{C10},\surv{S06},\surv{B13}                       & Beta Pic   &     24    &     19    &     29    &    51.5    &   -0.13    &    1.2    \\
\object{HIP\,93375}        & 19&01&06.00 & $-$28&42&50.00 & G1V    & \surv{C15}                                                        & AB Dor     &    149    &    100    &    180    &    58.8    &   -0.01    &    1.1    \\
\object{HIP\,95270}        & 19&22&58.90 & $-$54&32&17.00 & F6V    & \surv{C10},\surv{R13},\surv{B13}                                  & Beta Pic   &     24    &     19    &     29    &    51.8    &   -0.13    &    1.3    \\
\object{HIP\,96334}        & 19&35&09.72 & $-$69&58&32.09 & G3V    & \surv{C10},\surv{S06}                                             &            &    150    &     70    &    220    &    36.4    &    0.06    &    1.0    \\
\object{TYC\,5736-0649-1}  & 19&45&36.00 & $-$14&27&54.00 & G6V    & \surv{C15}                                                        &            &     40    &     10    &    100    &    86.4    &  \ldots    &    1.0    \\
\object{HIP\,97438}        & 19&48&15.45 & $+$59&25&22.45 & G0     & \surv{L07}                                                        & Cas?       &    450    &    250    &    700    &    27.7    &    0.08    &    1.2    \\
\object{HD\,189285}        & 19&59&24.10 & $-$04&32&06.00 & G5     & \surv{C15}                                                        & AB Dor     &    149    &    100    &    180    &    77.8    &   -0.01    &    1.0    \\
\object{HIP\,98470}        & 20&00&20.25 & $-$33&42&12.43 & F7V    & \surv{C15}                                                        & Her-Lyr?   &    150    &    100    &    300    &    21.2    &  \ldots    &    1.2    \\
\object{HIP\,99273}        & 20&09&05.22 & $-$26&13&26.63 & F5V    & \surv{C15},\surv{V12},\surv{R13},\surv{B13}                       & Beta Pic   &     24    &     19    &     29    &    52.2    &   -0.13    &    1.3    \\
\object{HIP\,101262}       & 20&31&32.07 & $+$33&46&33.13 & K5     & \surv{L07}                                                        &            &    700    &    400    &   1500    &    26.9    &    0.19    &    0.8    \\
\object{HIP\,102626}       & 20&47&45.02 & $-$36&35&40.83 & K0V    & \surv{L05},\surv{C10}                                             & Tuc?       &     45    &     10    &    100    &    52.2    &  \ldots    &    1.0    \\
\object{HIP\,104214}       & 21&06&53.94 & $+$38&44&57.89 & K5V    & \surv{B07},\surv{H10}                                             &            &   6000    &   2000    &   7000    &     3.5    &   -0.23    &    0.7    \\
\object{HIP\,104217}       & 21&06&55.26 & $+$38&44&31.40 & K7V    & \surv{H10}                                                        &            &   6000    &   2000    &   7000    &     3.5    &   -0.23    &    0.7    \\
\object{HIP\,104225}       & 21&06&56.39 & $+$69&40&28.55 & K0     & \surv{L07}                                                        &            &   5000    &   3000    &   8000    &    32.9    &   -0.30    &    0.8    \\
\object{TYC\,6351-0286-1}  & 21&13&05.30 & $-$17&29&13.00 & K6Ve   & \surv{B14}                                                        & AB Dor     &    149    &    100    &    200    &    38.5    &   -0.01    &    0.7    \\
\object{HIP\,105038}       & 21&16&32.47 & $+$09&23&37.77 & K2     & \surv{L07}                                                        &            &   1000    &    400    &   2500    &    16.4    &   -0.10    &    0.8    \\
\object{HIP\,105388}       & 21&20&50.00 & $-$53&02&03.00 & G7V    & \surv{L05}                                                        & Tuc/Hor    &     45    &     35    &     50    &    43.0    &   -0.06    &    0.9    \\
\object{HIP\,105612}       & 21&23&27.14 & $-$75&29&37.75 & G5V    & \surv{C15}                                                        &            &    600    &    300    &   1300    &    32.8    &  \ldots    &    1.0    \\
\object{HIP\,106231}       & 21&31&01.71 & $+$23&20&07.37 & K8     & \surv{L05},\surv{L07}                                             & AB Dor     &    149    &    100    &    180    &    24.8    &   -0.01    &    0.7    \\
\object{HIP\,107350}       & 21&44&31.33 & $+$14&46&18.98 & G0V    & \surv{L07},\surv{S06},\surv{B14}                                  & Her/Lyr?   &    300    &    200    &   1000    &    17.9    &    0.00    &    1.1    \\
\object{HIP\,107649}       & 21&48&15.75 & $-$47&18&13.01 & G2V    & \surv{K07}                                                        &            &   2500    &   1500    &   5000    &    16.0    &    0.00    &    1.0    \\
\object{HIP\,107947}       & 21&52&09.70 & $-$62&03&08.00 & F6V    & \surv{C10},\surv{S06},\surv{R13}                                  & Tuc/Hor    &     45    &     35    &     50    &    45.3    &   -0.06    &    1.2    \\
\object{HIP\,108156}       & 21&54&45.04 & $+$32&19&42.85 & K0     & \surv{L07}                                                        &            &   2000    &   1000    &   4000    &    20.0    &   -0.08    &    0.8    \\
\object{HIP\,108809}       & 22&02&32.96 & $-$32&08&01.49 & F6/F7V & \surv{L05},\surv{R13}                                             & UMa?       &    450    &    350    &    550    &    30.1    &   -0.02    &    1.2    \\
\object{HIP\,108870}       & 22&03&21.70 & $-$56&47&10.00 & K5V    & \surv{B07}                                                        &            &   2500    &   1500    &   4000    &     3.6    &   -0.20    &    0.7    \\
\object{V383\,Lac}         & 22&20&07.03 & $+$49&30&11.76 & K0     & \surv{L07},\surv{H10},\surv{S06}                                  &            &     80    &     40    &    160    &    35.4    &    0.28    &    0.9    \\
\object{HIP\,110996}       & 22&29&15.23 & $-$30&01&06.27 & K4V    & \surv{B07}                                                        &            &   4000    &   2000    &   8000    &    16.1    &    0.08    &    0.8    \\
\object{HIP\,111449}       & 22&34&41.64 & $-$20&42&29.56 & F7V    & \surv{L07},\surv{B14}                                             & Her/Lyr    &    250    &    200    &   1000    &    22.7    &    0.08    &    1.4    \\
\object{CP\,-72\,2713}     & 22&42&48.92 & $-$71&42&21.26 & K7V    & \surv{B13}                                                        & Beta Pic   &     24    &     19    &     29    &    36.0    &  \ldots    &    0.6    \\
\object{TYC\,8004-0083-1}  & 22&46&33.50 & $-$39&28&45.00 & G5V    & \surv{C15}                                                        &            &    150    &     80    &    200    &    74.9    &  \ldots    &    1.0    \\
\object{HIP\,113283}       & 22&56&24.05 & $-$31&33&56.00 & K4V    & \surv{L05},\surv{H10}                                             & Cas        &    440    &    400    &    480    &     7.6    &    0.00    &    0.7    \\
\object{HIP\,113579}       & 23&00&19.29 & $-$26&09&13.50 & G5V    & \surv{C10},\surv{S06}                                             & AB Dor     &    149    &    100    &    180    &    30.8    &   -0.01    &    1.0    \\
\object{HIP\,114530}       & 23&11&52.05 & $-$45&08&10.63 & G8V    & \surv{S06}                                                        & AB Dor     &    149    &    100    &    180    &    49.4    &   -0.01    &    1.0    \\
\object{HD\,219498}        & 23&16&05.02 & $+$22&10&34.82 & G5     & \surv{R13}                                                        &            &    270    &    150    &    400    &    62.1    &  \ldots    &    1.0    \\
\object{HIP\,115147}       & 23&19&26.63 & $+$79&00&12.67 & G9V    & \surv{L05},\surv{L07},\surv{H10}                                  &            &     52    &     30    &     80    &    19.2    &   -0.03    &    0.8    \\
\object{HIP\,115162}       & 23&19&39.56 & $+$42&15&09.82 & G0     & \surv{S06},\surv{B14}                                             & AB Dor     &    149    &    100    &    180    &    50.2    &   -0.01    &    1.0    \\
\object{FTYC\,9338-2016-1} & 23&21&52.50 & $-$69&42&12.00 & G8V    & \surv{C15}                                                        &            &     45    &     30    &    100    &    99.6    &  \ldots    &    1.0    \\
\object{TYC\,9529-0340-1}  & 23&27&49.40 & $-$86&13&19.00 & G8IV   & \surv{C15}                                                        & Tuc/Hor    &     45    &     35    &     50    &    68.8    &   -0.06    &    1.0    \\
\object{TYC\,9339-2158-1}  & 23&31&00.80 & $-$69&05&11.00 & K3V    & \surv{C15}                                                        &            &    350    &    200    &    500    &    30.6    &  \ldots    &    0.8    \\
\object{TYC\,6406-0180-1}  & 23&33&06.10 & $-$17&54&42.00 & K0V    & \surv{C15}                                                        &            &    250    &    150    &   4000    &    58.0    &  \ldots    &    0.9    \\
\object{HIP\,116431}       & 23&35&36.15 & $+$08&22&57.43 & F0     & \surv{V12},\surv{R13}                                             &            &    150    &     50    &   2000    &    68.4    &   -0.09    &    1.4    \\
\object{HIP\,116910}       & 23&41&54.30 & $-$35&58&40.00 & G8V    & \surv{C15}                                                        & AB Dor     &    149    &    100    &    180    &    63.7    &   -0.01    &    1.0    \\
\object{HIP\,118008}       & 23&56&10.67 & $-$39&03&08.40 & K3V    & \surv{B07},\surv{C10},\surv{S06}                                  & AB Dor     &    149    &    100    &    180    &    22.0    &   -0.01    &    0.8    \\
\end{longtable}
\tablefoot{\tablefoottext{a}{M05: \citet{masciadri2005}, L05: \citet{lowrance2005}, S06: Song et al. (unpublished), K07: \citet{kasper2007}, B07: \citet{biller2007}, L07: \citet{lafreniere2007}, C10: \citet{chauvin2010}, H10: \citet{heinze2010}, V12: \citet{vigan2012}, R13: \citet{rameau2013}, B13: \citet{biller2013}, B14: \citet{brandt2014}, C15: \citet{chauvin2015}.} \tablefoottext{b}{AB Dor: AB Doradus, Arg: Argus, Beta Pic: Beta Pictoris, Car: Carina, Cas: Castor, Col: Columba, Her/Lyr: Hercules/Lyrae, LA: Local association, LCC: Lower Centaurus Crux, Tuc/Hor: Tucana/Horlogium, UCL: Upper Centaurus Lupus, UMa: Ursa Major, US: Upper Scorpius.}}

\end{landscape}

\section{Impact of stellar ages on the frequency estimation}
\label{sec:app_freq}

The values of the probability density function of $f$ in Tables~\ref{tab:freq} and \ref{tab:freq_pop} are provided for the best age estimates of the stars. In Tables~\ref{tab:app_freq} and \ref{tab:app_freq_pop} we also provide the values for the minimum and maximum stellar ages. The assumptions are the same as detailed in Sect.~\ref{sec:frequency_estimation} and Sect.~\ref{sec:synth_pop_mess} respectively.

\begin{table}[!h]
    \caption[]{Fraction of stars that host a planetary system, assuming linear flat mass and semi-major axis distributions.}
    \label{tab:app_freq}
    \centering
    \begin{tabular}{cccccc}
    \hline \hline
    SMA     & Mass               & $N_{det}$\tablefootmark{a} & $F_{best}$\tablefootmark{b} & \multicolumn{2}{c}{$[F_{min}, F_{max}]$\tablefootmark{c}} \\
    (AU)    & ($M_\mathrm{Jup}$) &                            & (\%)                        & CL=68\%      & CL=95\%       \\
    \hline
    \multicolumn{6}{c}{Minimum age} \\
    \hline
     5--500 &   5--75            & 3                          & 2.40                        & [1.70, 4.60] & [0.90, 6.80]  \\
     5--500 &   5--14            & 1                          & 0.90                        & [0.70, 2.95] & [0.25, 5.00]  \\
    20--300 & 0.5--75            & 3                          & 2.05                        & [1.45, 3.90] & [0.75, 5.80]  \\
    20--300 & 0.5--14            & 1                          & 1.00                        & [0.70, 3.15] & [0.25, 5.35]  \\
    \hline
    \multicolumn{6}{c}{Best age} \\
    \hline
     5--500 &   5--75            & 3                          & 2.45                        & [1.70, 4.70] & [0.90, 6.95]  \\
     5--500 &   5--14            & 1                          & 1.00                        & [0.75, 3.30] & [0.25, 5.55]  \\
    20--300 & 0.5--75            & 3                          & 2.10                        & [1.50, 4.05] & [0.80, 5.95]  \\
    20--300 & 0.5--14            & 1                          & 1.15                        & [0.85, 3.65] & [0.30, 6.20]  \\
    \hline
    \multicolumn{6}{c}{Maximum age} \\
    \hline
     5--500 &   5--75            & 3                          & 2.50                        & [1.75, 4.85] & [0.95, 7.15]  \\
     5--500 &   5--14            & 1                          & 1.15                        & [0.85, 3.70] & [0.30, 6.20]  \\
    20--300 & 0.5--75            & 3                          & 2.15                        & [1.55, 4.20] & [0.80, 6.20]  \\
    20--300 & 0.5--14            & 1                          & 1.35                        & [0.95, 4.25] & [0.35, 7.20]  \\
    \hline
    \end{tabular} 
    \tablefoot{\tablefoottext{a}{Number of detections in the considered mass and semi-major axis (SMA) range.} \tablefoottext{b}{Best value of the planet frequency compatible with the observations.} \tablefoottext{c}{Minimum and maximum values of the frequency compatible with the results for a given confidence level (CL).}}
\end{table}

\begin{table}[!h]
    \caption[]{Fraction of stars that host a planetary system based on the GI populations described in Sect.~\ref{sec:planet_population_synthesis_models}.}
    \label{tab:app_freq_pop}
    \centering
    \begin{tabular}{cccccc}
    \hline \hline
    Mass               & Scattering & $N_{det}$\tablefootmark{a} & $F_{best}$\tablefootmark{b} & \multicolumn{2}{c}{$[F_{min}, F_{max}]$\tablefootmark{c}} \\
    ($M_\mathrm{Jup}$) &            &                            & (\%)                        & CL=68\%      & CL=95\%       \\
    \hline
    \multicolumn{6}{c}{Minimum age} \\
    \hline
    $\le$75           & No         & 3                          & 2.40                         & [1.70, 4.60] & [0.90 6.80] \\
    $\le$14           & No         & 1                          & 1.25                         & [0.95, 4.10] & [0.35 6.90] \\
    \hline
    $\le$75           & Yes        & 3                          & 2.70                         & [1.90, 5.20] & [1.00 7.70] \\
    $\le$14           & Yes        & 1                          & 1.50                         & [1.10, 4.90] & [0.40 8.25] \\
    \hline
    \multicolumn{6}{c}{Best age} \\
    \hline
    $\le$75           & No         & 3                          & 2.65                        & [1.85, 5.10] & [1.00, 7.50] \\
    $\le$14           & No         & 1                          & 1.60                        & [1.15, 5.20] & [0.40, 8.70] \\
    \hline
    $\le$75           & Yes        & 3                          & 3.00                        & [2.15, 5.85] & [1.10, 8.60] \\
    $\le$14           & Yes        & 1                          & 1.95                        & [1.40, 6.20] & [0.50, 10.45] \\
    \hline
    \multicolumn{6}{c}{Maximum age} \\
    \hline
    $\le$75           & No         & 3                          & 3.00                        & [2.10, 5.75] & [1.10 8.50] \\
    $\le$14           & No         & 1                          & 2.05                        & [1.50, 6.60] & [0.50 11.10] \\
    \hline
    $\le$75           & Yes        & 3                          & 3.45                        & [2.40, 6.60] & [1.25 9.75] \\
    $\le$14           & Yes        & 1                          & 2.45                        & [1.75, 7.95] & [0.60 13.30] \\
    \hline
    \end{tabular} 
    \tablefoot{\tablefoottext{a}{Number of detections in the considered mass range.} \tablefoottext{b}{Best value of the planet frequency compatible with the observations.} \tablefoottext{c}{Minimum and maximum values of the frequency compatible with the results for a given confidence level (CL).}}
\end{table}

\end{document}